\documentclass[10pt,a4paper,footsepline,headsepline,headinclude,twocolumn]{scrartcl}

\pdfoutput=1

\usepackage{setspace}

\usepackage{amssymb}
\usepackage{amsmath}
\usepackage{amsthm}
\usepackage{amsfonts}
\usepackage{mathrsfs}

\usepackage{a4wide}
\usepackage{scrpage2}

\usepackage[english]{babel}
\usepackage[T1]{fontenc}
\usepackage[utf8x]{inputenc}

\setfootsepline{.5pt}

\usepackage{graphicx}
\usepackage{color}

\usepackage[hangindent=0pt,singlelinecheck=on,format=plain,font=scriptsize]{caption}
\usepackage[font=scriptsize,labelformat=parens]{subcaption}

\usepackage{float}
\floatstyle{boxed}
\restylefloat{figure}

\usepackage[algoruled,linesnumbered]{algorithm2e}
\makeatletter
\newcommand{\removelatexerror}{\let\@latex@error\@gobble}
\makeatother

\newcommand{\myalgorithm}{
  \begingroup
  \removelatexerror
  \begin{algorithm*}[H]
    \label{alg1}
    \RestyleAlgo{ruled}
    \SetAlgoLined
    \LinesNumbered
    \SetKwInOut{Input}{input}
    \SetKwInOut{Output}{output}
    
    \Input{Surface meshes $\mathcal{S},\mathcal{T}$ and elasticity
      parameters}

    \Output{Deformations $\Phi,\vphi^h$, deformed meshes
      $\Phi(\mathcal{S}),\vphi^h(\mathcal{B}^h)$ and forces $\tilde
      G_B$}

    \Begin{

      Create a tetrahedralization $\mathcal{B}^h$ of $\mathcal{S}$\;
    
      Project the nodes of $\mathcal{S}$ onto the mesh
      $\partial\mathcal{B}^h$ and compute the smoothing map
      $T:C^0(\partial\mathcal{B}^h,\real^3)\rightarrow
      C^0(\mathcal{S},\real^3)$ described in~(\ref{eq:4-2})\;
      
      Precompute a local orthonormal coordinate system at each point
      $q\in\mathcal{T}$ to set up a local metric as described
      in~(\ref{eq:3-12})\;
      
      Set $\vphi_0^h=\id$, $\Phi=\id$, $E=0$ and $k=1$\;
      
      \While{$k\leq K_{\text{max}}$ and $E$ is not stationary}{
        
        Compute an initial guess of correspondences
        $\Phi^{h,\ast}:\Phi(\mathcal{S})\rightarrow\mathcal{T}$\;
      
        For each point $\Phi_{i}^\ast=\Phi^{h,\ast}(x_i)$ chose the
        corresponding metric $Q_{\Phi_{i}}^\ast$ as
        in~(\ref{eq:3-12})\;
        
        Set $l=1$;

        \While{$l\leq L_{\text{max}}$}{
          
          Compute the first order approximation~(\ref{eq:3-4}), i.e.,
          the zero order term and the tangent stiffness matrix at
          $\vphi_0^h$\;
          
          Solve the local optimization problem~(\ref{eq:4-3}) with
          initial deformation $\vphi_0^h$ in~(\ref{eq:4-4}) to obtain
          $\vphi_{B,k}^h,\Phi,\tilde G_B$\;
          
          Use the optimized boundary conditions
          $\vphi_{B,k}^h:\partial\mathcal{B}^h\rightarrow\real^3$ to
          solve for the full deformation
          $\vphi_k^h:\mathcal{B}^h\rightarrow\real^3$ by means of a
          Newton-Raphson solver\;
          
          Update $\vphi_0^h=\vphi_k^h$ and $l=l+1$\;
          
        }
        
        Save the stored energy $E=\vphi_k^hG_B$ of the deformed body
        and optionally various other elasticity measures (e.g.,
        strain, stress, etc.)\;
        
        Increase the spring constant in~(\ref{eq:4-3}) and possibly
        decrease the ratio $\lambda_n/\lambda_t$ in~(\ref{eq:3-12})\;

        Update $k=k+1$\;

      }
      
      Return\;

    }
    \caption{Sketch of the algorithm}
\end{algorithm*}
\endgroup}

\usepackage{dblfloatfix}
\usepackage{fixltx2e}

\usepackage{lipsum} 
\usepackage{authblk}

\newcommand {\real}{\mathbb{R}}

\newcommand {\tor}{\mathbb{S}}
\newcommand {\id}{\mathbb{I}}

\newcommand {\eps} {\varepsilon}
\newcommand {\vphi} {\varphi}

\DeclareMathOperator*{\minimize}{minimize}

\DeclareMathOperator{\DIV}{div}
\DeclareMathOperator{\drm}{d}

\newcommand{\diag}[1]{\operatorname{diag}(#1)} 
\newcommand{\tr}[1]{\operatorname{tr}#1} 
\newcommand{\cof}[1]{\operatorname{Cof}#1} 

\newcommand{\dual}[2]{\left\langle #1, #2\right\rangle}
\newcommand{\norm}[2]{\left\| #1 \right\|_{#2}}

\lohead{\today}
\rohead{Nonlinear Force Optimization for Shape Matching}
\rofoot{\pagemark} \lofoot{} 

\begin{document}

\thispagestyle{empty}

\vspace{1.0cm}

\begin{center}
  \LARGE \bf{A Hyperelastic Two-Scale Optimization Model for Shape
    Matching}
\end{center}
 
\vspace*{1cm}

\begin{center}
  Konrad Simon\textsuperscript{1}, Sameer Sheorey\textsuperscript{2},
  David W. Jacobs\textsuperscript{3} and Ronen Basri\textsuperscript{1}
\end{center}

\vskip 1cm  

\begin{center}
  \small {	\textsuperscript{1}~Department of Computer Science and Applied Mathematics, \\
  	The Weizmann Institute of Science, Rehovot 76100, Israel \\[0.5cm]
	\textsuperscript{2}~UtopiaCompression Corporation, \\
                  Los Angeles, CA 90064, USA \\[0.5cm]
	\textsuperscript{3}~Department of Computer Science, University of Maryland, \\
	College Park, MD 20742, USA
	}
\end{center}

\vspace{1cm}

\thispagestyle{empty}

\begin{abstract}
  \textbf{Abstract.} \small We suggest a novel shape matching algorithm for
  three-dimensional surface meshes of disk or sphere topology. The method is
  based on the physical theory of nonlinear elasticity and can hence handle
  large rotations and deformations. Deformation boundary conditions that
  supplement the underlying equations are usually unknown. Given an initial
  guess, these are optimized such that the mechanical boundary forces that are
  responsible for the deformation are of a simple nature. We show a heuristic
  way to approximate the nonlinear optimization problem by a sequence of
  convex problems using finite elements. The deformation cost, i.e., the
  forces, is measured on a coarse scale while ICP-like matching is done on the
  fine scale. We demonstrate the plausibility of our algorithm on examples
  taken from different datasets.
\end{abstract}

\vspace{0.7cm}


\setcounter{page}{1}

\section{Introduction}
	\subsection{Motivation}

Shape matching is a difficult but nonetheless important problem in computer
vision, medical imaging, and computer graphics. It is the key ingredient for
recognition, retrieval, alignment of scanned data, information transfer, shape
interpolation, statistical shape modeling, space-time reconstruction and more.

This work aims to contribute a novel framework to solve for unknown shape
deformations of a pair of two-dimensional surfaces (source and target) in
three dimensions. Our method is built on the observation that in applications
surfaces often represent the boundaries of actual physical entities, i.e.,
surfaces of elastic bodies. Shape change can therefore be explained by means
of forces acting on the specific elastic body. Their magnitude can be
interpreted as a measure of how ``difficult'' it is to achieve a certain shape
change. Elastic models are, we believe, well suited to give valuable
information about shape changes among different objects. In particular, we
believe, that sparsity of forces is a good prior for explaining an observed
deformation within a semantic class of objects, e.g., if we were to compare
two different shape articulations.

From a mathematical point of view shape matching is an ill-posed inverse
problem and is usually tackled in a heuristic manner. It is computationally
challenging since the search space for correspondences is usually
large. Hence, there is a need to explore this search space in a reasonable
manner.

\subsection{Our Method}

In~\cite{Sederberg} the authors intuitively describe the desired shape
blending algorithm as one which requires the least work to deform through
bending and stretching. Many shape matching algorithms are designed in this
spirit as one usually seeks ``maximum alignment by means of minimal cost'' in
an optimization framework. Our work uses the physical theory of nonlinear
elasticity and is based on the same principle as will be elaborated in the
following.

In this work we suggest a method for aligning two surface meshes (manifolds)
embedded in three-dimensional space. We use the observation that surface
meshes often represent (parts of) the boundaries of actual physical
bodies. Hence, the cost of a deformation of a surface can be measured by
forces acting on the boundary of the solid body described by the surface.

We believe that in many realistic scenarios even complicated deformations take
place due to or can be explained by means of surprisingly simple
forces. Examples are the change of a pose of an animal or a human where forces
act mainly only on the articulated parts, i.e., the acting forces can be
regarded as sparse. We furthermore anticipate that, given two semantically
similar objects, seeking a deformation that is subject to sparse and isotropic
surface forces is a reasonable strategy to match them. These forces could then
be used to intuitively compare two different deformations and to assess their
severity.

The underlying equations of nonlinear elasticity that we employ form a system
of nonlinear partial differential equations (PDEs) and are subject to boundary
conditions. These can be given as deformations and/or forces prescribed on the
boundary of the body. Given two shapes boundary conditions of either type are
usually unknown but one can often give an initial guess of boundary
correspondences, i.e., of the boundary deformation. Such an initial guess can
then be optimized according to our prior on the forces to obtain a ``better''
set of boundary correspondences that supplement the PDE. The initial guess
plays the role of a spring force between source and target shape and a penalty
needs to be paid for deviation. The optimized boundary correspondences can
then be used to solve the underlying elasticity PDE using a common
Newton-Raphson scheme to update the full volumetric deformation of the body
and the entire procedure can be iterated until a good match is found.

Using the popular finite element method (FEM) we show how to connect the
boundary deformation to the boundary forces by means of a splitting of the
tangent stiffness matrix, localized around a certain deformation. This
so-called condensation is necessary to reduce the optimization to the boundary
of the body and has the advantage that it lowers the dimension of the
optimization significantly. Furthermore, the condensation allows us to
formulate the problem as a sequence of (convex) second order cone problems.

As elaborated above we aim to match surfaces represented by triangular meshes
but the FEM works on volumetric meshes. Hence, we create a tetrahedral mesh of
the body represented by the triangular mesh. The surface meshes that we match
in our examples have between 9000 to 55000 triangles and constructing a
tetrahedral mesh that resembles the surface density of meshes would result in
a computationally very large problem. Therefore, we create a coarse
tetrahedral approximation of the surface. Every point on the surface mesh can
then be coupled to the coarse boundary mesh of the volumetric mesh. A
deformation of the coarse boundary then induces a deformation of the surface.
This way we can prescribe a matching (in particular the initial guess) on the
surface mesh while measuring the deformation cost on the coarser volumetric
mesh.

\subsection{Previous Work}

\subsubsection{Overview}
An optimization based shape matching algorithm usually consists of two
ingredients. First, one needs a regularizer that determines (the preference
of) the class of deformations in consideration. Well investigated is the field
of rigid matching, i.e., the admissible transformations are translations and
rotations~\cite{Aiger,Gelfand,Rusinkiewicz}. This is a common scenario if one
is given a range scan of rigid objects that can be aligned with a single rigid
transform. For non-rigid matching common regularizers that are used in the
literature are the popular As-Rigid-As-Possible (ARAP) functional~\cite{Alexa}
and the As-Affine-As-Possible (AAAP) functional~\cite{Kovalsky,Li}. Both are
often combined in one objective function. For a set of correspondences AAAP
measures the deviation from a global affine transform and hence favors smooth
deformations. ARAP locally measures the deviation from a rigid transform and
is related to elasticity since it mimics mechanical stiffness. These methods
work on tetrahedral meshes, i.e., they need volumetric meshes that represent
the shape. In the context of surfaces recent ARAP-like methods have been
introduced in~\cite{Sorkine,LeviGotsman}.

Elastically deformable models in graphics have been pioneered in the end of
the 1980s in~\cite{Terzopoulos}. Generally speaking, these models use
characterizing properties of geometric objects (curves, surface, volumes) to
define functionals that penalize deviation from them. A curve in three
dimensions, for example, is fully characterized (up to rigid motion) by two
parameters, its curvature and its torsion and hence a motion of the curve that
changes one of them is considered a deformation of the curve. For a volume it
is the local isotropic change of lengths that characterizes a deformation (up
to rigid motion). This principle is reflected in the physical theory of
(nonlinear) elasticity~\cite{Antman,Ciarlet}. In the context of shape matching
deformation models that borrow ideas from elasticity can be found
in~\cite{Ferrant}, where the authors use linear elasticity for the
registration of three-dimensional medical images, as well as in~\cite{Choi}
for fitting solid meshes to animated surfaces. In~\cite{Chui} a thin-plate
spline regularizer is used for point matching. Non-linear elasticity is
employed in~\cite{RabbitWeissChristensenMiller} for medical imaging and
in~\cite{Savran} where two-dimensional elasticity is employed for
three-dimensional surface matching via mesh parametrization. However, there is
a vast amount of (physical) deformation methods and we do not aim at
discussing them in detail. An overview of methods used in graphics can be
found in~\cite{Nealen} and techniques used in vision and medical imaging
in~\cite{Holden}.

\vspace{0.5cm}

The second ingredient for a shape matching method is a data term that drives
the deformation and/or describes dissimilarity between source and target
shape. One way is to randomly sample candidate correspondences and to choose a
deformation that best aligns the data. These RANSAC methods, which were first
introduced in~\cite{Fischler}, are practical for low-dimensional mapping
spaces. Recently they were extended to deal with deformation spaces of higher
dimension~\cite{Lipman2}.

Another popular group of methods comprises variants of the famous iterative
closest point (ICP) algorithm, first introduced in~\cite{Besl,Chen}. This
iterative approach was first used for rigid matching and is well suited for
various representations of geometric data. It is based on the computation of
closest points between source and target. Since then efficient methods have
been developed that use different sampling of correspondences, different
weighting schemes that determine the confidence of a candidate match etc.,
see~\cite{Rusinkiewicz,Choi,Kovalsky}. Recently, ICP has been applied as well
in the context of non-rigid matching~\cite{Allen,Bronstein,Brown,Huang}. Some
of these papers deal with deformations of significant magnitude.

Techniques like multi-dimensional scaling (MDS) and generalized
multi-dimensional scaling (GMDS) aim to find correspondences by regarding the
shapes as Riemannian manifolds and try to find an embedding in a common metric
space~\cite{Bronstein,TOSCA}. Correspondences are then found in the common
embedding space (which can even be a high-dimensional Riemannian manifold in
the case of GMDS) rather than between the shapes themselves. This way ICP can
be regarded as an instance of the Euclidean isometric matching problem. MDS
and GMDS are elegant methods but computationally costly.

We note, however, that the vast literature and number of techniques can be
classified also by other criteria. An excellent overview of the field and of
classification criteria can be found in the survey paper~\cite{Kaick}.

\subsubsection{Comparison}
This work extends our work in~\cite{Simon} to three dimensions and large
deformations. By favoring sparse forces as explanations for elastic
deformations we use the same prior that we used in~\cite{Simon}. There we deal
with small planar deformations. The forces there are invariant under
infinitesimal rotations only, due to the use of linear elasticity and are, in
particular, not invariant under general rotations.

In~\cite{Choi} the authors employ an enhanced version of linear elasticity by
the use of rotation compensation, similar to co-rotated linear
elasticity. This way they can deal with large rotations but this requires a
large number of low dimensional singular value decompositions in each
iteration, whereas we solely update the tangent stiffness matrix computed by
the FEM. Another difference is that in our method the optimized boundary
forces can be made rotation invariant once a matching deformation is found,
and it is theoretically suitable for very large deformations.

The splitting of the tangent stiffness matrix that we employ to connect
boundary forces and boundary displacements was previously used
in~\cite{Bro-Nielsen} in the context of surgery simulation and in~\cite{Simon}
for matching. The technique that we propose generalizes this idea to the
nonlinear case and allows us to reduce the (nonlinear) matching problem to
solving a sequence of convex problems that we formulate as second order cone
problems while simultaneously reducing the computational complexity.

In our algorithm, the computational complexity, i.e., the degrees of freedom
to be optimized, is furthermore reduced by measuring the deformation cost,
i.e., the elastic forces, on a (coarse) tetrahedral mesh that is needed by the
FEM. The corresponding deformation is then extended to the (fine) triangular
surface mesh. Similar ideas have used in~\cite{Kovalsky,Li,Sumner}.
Furthermore, changing the coarseness of the tetrahedral mesh gives the user
control over the computational complexity of the FEM part of the model while
the coarseness of the surface mesh is fixed.

The matching and the computation of the initial guess is done on the fine
scale of the surfaces. We project the source points onto the target favoring
similar points (descriptor aided) during the first few iterations and then
switch to Euclidean projection once a better alignment is reached. This way
one can regard our method as a two-scale version of a non-rigid ICP-like
algorithm since the matching is done on the fine scale while the cost, i.e.,
the boundary forces, is measured on the coarse scale of the volumetric mesh.

\vspace{0.5cm}

The paper is organized as follows. In Section~\ref{s-2} we give a concise
overview of the model of a hyperelastic body and describe a nonlinear FEM that
is used in the optimization procedure. The optimization is described in detail
in Section~\ref{s-3}. Experiments, implementation details and a description of
the two-scale algorithm are given in Section~\ref{s-4}. Section~\ref{s-5}
concludes with a discussion.

\section{Model and Numerical Methods}\label{s-2}
	\subsection{Hyperelasticity}\label{s-2-1}

The purpose of this section is to introduce the reader into the most important
principles of nonlinear elasticity theory that are relevant in the sequel. As
mentioned, our model optimizes elastic forces and hence it is important to
understand what forces we refer to and how they act. Nonlinear elasticity is a
well developed and complex field. Mathematically oriented introductions can be
found in~\cite{Antman,Ciarlet,Braess} and an engineering oriented introduction
in~\cite{Dhondt}.

Roughly speaking an elastic body is an open connected subset $\mathcal{B}
\subset \real^3$ that reacts to applied forces with a deformation. It returns
to its rest state (also called reference configuration) after the forces are
removed and does not memorize previous deformations. The main equations of
nonlinear elasticity describe a balance between the applied forces that cause
the deformation and the internal force distribution inside the deformed body
(stress) in the spirit of Newton's second law. We will elaborate on this in
the following.

\vspace{0.5cm}

\textbf{Stress, Forces and Equilibrium.} We assume that we are given an
elastic body in its rest pose $\mathcal{B}\subset\real^3$, i.e., the rest pose
is the volume that is occupied by the body when no forces are applied to it. A
deformation of the body is described by a locally injective and smooth vector
field $\Phi:\mathcal{B}\rightarrow \real^3$, i.e., any point $x\in
\mathcal{B}$ in the reference configuration has a corresponding point
$x^\Phi:=\Phi(x)\in \mathcal{B}^\Phi:=\Phi(\mathcal{B})$ in the deformed
configuration. Any deformation should satisfy $\det (\nabla\Phi)>0$, i.e.,
subsets of $\mathcal{B}$ of positive measure are mapped to subsets of
$\mathcal{B}^\Phi$ of positive measure.

\begin{figure*}[t!]
  \centering
  \includegraphics[width=0.95\textwidth]{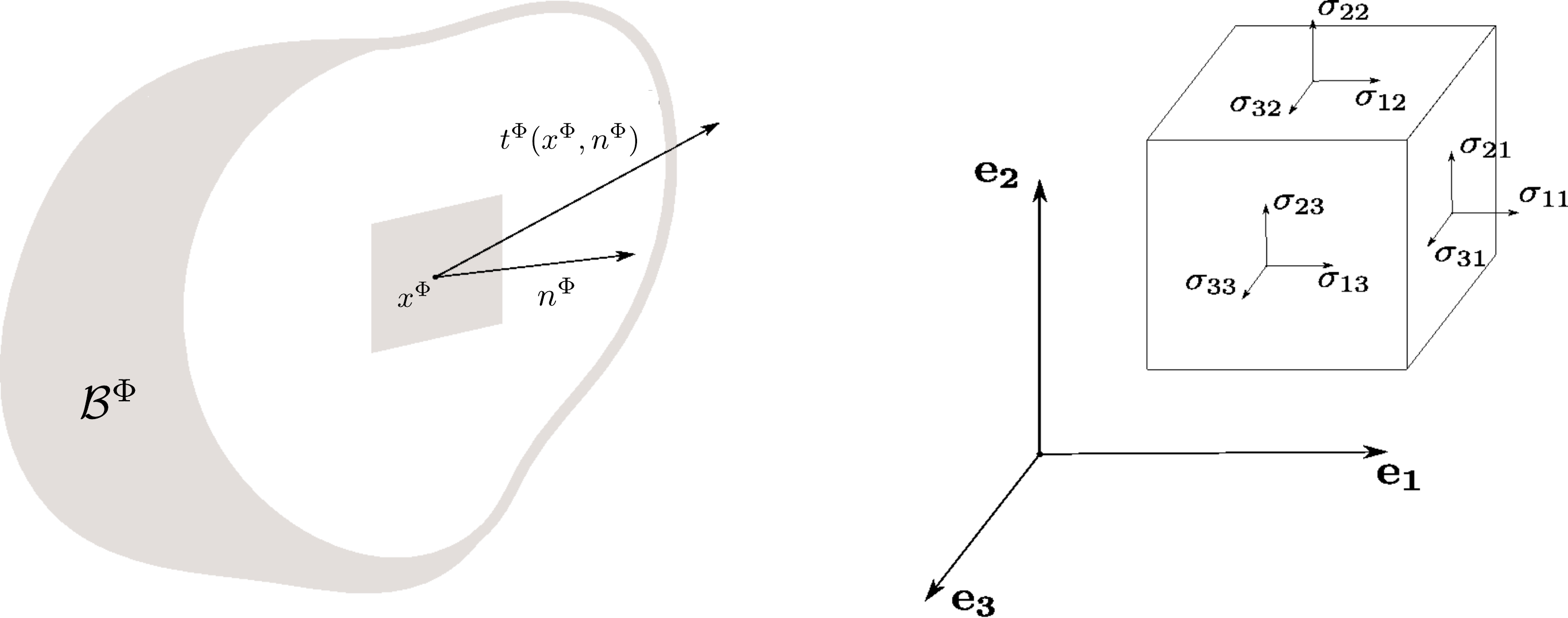}
  \caption{Left: an illustration of a surface force $t^\Phi(x^\Phi,n^\Phi)$ at
    some point $x^\Phi$ in an arbitrary cross section with normal $n^\Phi$ of
    a body $\mathcal{B}^\Phi$.  Right: illustration of the three components of
    the Cauchy stress tensor $\sigma=T^\Phi$ with respect to the canonical
    coordinate planes. The normal stress is orthogonal to the considered plane
    shear stresses lie within the plane.}
  \label{fig-stress}
\end{figure*}

Suppose now that the rest state $\mathcal{B}$ is subject to forces that
describe the action of the outside world on it. These can either be volumetric
forces $f^\Phi$, i.e., they are measured by per unit volume of the deformed
configuration, or they can be surface forces $g^\Phi$ measured per unit
surface area acting on $\partial \mathcal{B}^\Phi$. Examples for volumetric
forces are gravity or electric field forces, whereas examples for surface
forces are pressure or spring forces. The resulting deformation $\Phi$ induces
an internal force distribution $t^\Phi:\mathcal{B}^\Phi\times
\tor^2\rightarrow\real^3$ inside $\mathcal{B}^\Phi$ that balances the external
forces, the so-called Cauchy traction vector. This vector field depends on two
arguments since it measures the force per unit area acting through any cross
section of $\mathcal{B}^\Phi$ described by its unit normal $n^\Phi\in \tor^2$
at $x^\Phi\in \mathcal{B}^\Phi$. This can be expressed through a linear
relation $t(x^\Phi,n^\Phi)=T^\Phi(x^\Phi)n^\Phi$ where
$T^\Phi:\mathcal{B}^\Phi\rightarrow \real^{3\times 3}$ is the Cauchy stress
tensor. Its rows represent the three tractions (normal and shear stresses) in
three coordinate planes which are usually the three orthogonal canonical
planes, see Figure~\ref{fig-stress} for an illustration. In particular, if
$x^\Phi\in\partial \mathcal{B}^\Phi$ and $n^\Phi\in \tor^2$ is its
corresponding outward unit normal, then $t^\Phi(x^\Phi,n^\Phi)$ is the force
$g^\Phi$ acting on the boundary of $\mathcal{B}^\Phi$ at this specific
point. Boundary forces will be of importance in the sequel for the formulation
of our optimization problem.

The basic equations governing elasticity theory can now be formulated as
follows (Cauchy principle): Let the volumetric forces be denoted by
$f^\Phi:\mathcal{B}^\Phi\rightarrow\real^3$ and let $\mathcal{V}\subset
\mathcal{B}^\Phi$ be an arbitrary subvolume. Then
\begin{equation}
  \label{eq:2-1}
  \int_{\partial \mathcal{V}} t^\Phi(x^\Phi,n^\Phi) \drm S^\Phi + \int_\mathcal{V} f^\Phi(x^\Phi) \drm x^\Phi = 0 \:.
\end{equation}
This equilibrium of external forces and internal stress is an expression of
Newton's second law and holds in the deformed configuration
$\mathcal{B}^\Phi$. A similar principle for the angular moments shows in
addition that $T^\Phi$ must be symmetric. Using the above and the divergence
theorem we get
\begin{equation}
  \label{eq:2-2}
  \int_\mathcal{V} \left( \DIV^\Phi T^\Phi(x^\Phi) + f^\Phi(x^\Phi) \right) \drm x^\Phi = 0 \:.
\end{equation}
Since this holds for all subvolumes $\mathcal{V}$ the equilibrium equations can
be written in differential form as
\begin{equation}
  \label{eq:2-3}
  \begin{split}
    -\DIV^\Phi T^\Phi(x^\Phi) & = f^\Phi(x^\Phi) \:, \quad x^\Phi\in \mathcal{B}^\Phi
    \:, \\
    T^\Phi(x^\Phi)n^\Phi & = g^\Phi(x^\Phi) \:, \quad x^\Phi\in \partial \mathcal{B}^\Phi
  \end{split}
\end{equation}
where $g^\Phi$ is a boundary force as described above.

Unfortunately, this is not practical yet since equation~(\ref{eq:2-3}) holds
in the deformed configuration which is unknown. This can be remedied by a
pullback to the reference configuration $\mathcal{B}$. The main tool for this
is the so called Piola transform given by:
\begin{equation}
  \label{eq:2-4}
  T(x) = \det(\nabla\Phi(x)) T^\Phi(x^\Phi) \nabla\Phi(x)^{-T}\:.
\end{equation}
It describes a transition from the Eulerian variable $x^\Phi=\Phi(x)$ to the
Lagrangian (reference) variable $x$. The quantity $T$ is called the first
Piola-Kirchhoff stress tensor. It measures stress on the deformed
configuration per unit area of the undeformed configuration and it is not
symmetric. The Piola transform also transforms operators and forces
in~(\ref{eq:2-3}) so that the equilibrium equations in the reference
configuration $\mathcal{B}$ take the form
\begin{equation}
  \label{eq:2-5}
  \begin{split}
    -\DIV T(x) & = f(x) \:, \quad x\in \mathcal{B} \:, \\
    T(x)n & = g(x) \:, \quad x\in \partial \mathcal{B} \:.
  \end{split}
\end{equation}
The exact relations between the forces is given by
\begin{equation}
  \label{eq:2-5a}
  \begin{split}
    f(x) & = ( \det\nabla\Phi(x) )f^\Phi(x^\Phi) \:, \\
    g(x) & = ( \det\nabla\Phi(x) )\norm{\nabla\Phi(x)^{-T}n(x)}{}g^\Phi(x^\Phi) \:. \\
  \end{split}
\end{equation}
This means that the transformed forces in general depend on the deformation
$\Phi$ and its gradient even though this might not be the case for $f^\Phi$
and $g^\Phi$. The interested reader is referred to~\cite{Ciarlet}. The second
Piola-Kirchhoff stress tensor, given by
\begin{equation}
  \label{eq:2-6}
  \Sigma(x) = \nabla\Phi(x)^{-1} T(x) \:,
\end{equation}
is symmetric and measures stress on the undeformed configuration per
unit area of the undeformed configuration. Equation~(\ref{eq:2-5}) can
be rewritten in terms of $\Sigma$ as
\begin{equation}
  \label{eq:2-7}
  \begin{split}
    -\DIV \left( \nabla\Phi(x)\Sigma(x) \right) & = f(x) \:, \quad x\in \mathcal{B} \:, \\
    \Sigma(x)n & = \tilde g(x) \:, \quad x\in \partial \mathcal{B}
  \end{split}
\end{equation}
where $\tilde g(x) = \nabla\Phi(x)^{-1}g(x)$. However, we will
use~(\ref{eq:2-5}) as a basis for discretization and optimization.

\vspace{0.5cm}

\textbf{Constitutive Laws.} The above given equations describe equilibria of
external and internal forces but they do not take into account the different
reactions of materials to forces. Obviously, steel reacts with a different
deformation than rubber when exposed to the same external forces. Or,
conversely, rubber and steel undergoing the same deformation is caused by
different forces. These material specific relations between force and induced
deformation are modeled by so-called constitutive laws and shall be described
briefly in the following.

Roughly speaking, a material is called elastic if the Cauchy stress $T^\Phi$
can be written as a function of $x\in \mathcal{B}$ and $\nabla\Phi$. With
respect to~(\ref{eq:2-4}) and~(\ref{eq:2-6}) we can call a material elastic if
its first and second Piola-Kirchhoff stresses are functions of $x\in
\mathcal{B}$ and $\nabla\Phi$ only. Such a function is called a response
function. This definition allows a large class of response functions and can
be restricted further by making reasonable physical assumptions. A common
physical principle is the principle of frame indifference meaning that any
observed quantity is independent of the observer. Furthermore, one can assume
the independence of the material response on $x\in \mathcal{B}$. This is
called homogeneity. Another restriction of the class of admissible response
functions is the isotropy of the material which essentially means that the
materials' response at any point $x\in \mathcal{B}$ is the same and is
independent of the direction in which the force is applied. Steel is an
isotropic material in contrast to wood which deforms differently in the
direction of its fibers than orthogonal to them. Combining all these
assumptions one can show that the Cauchy stress and the second Piola-Kirchhoff
stress are of the form
\begin{equation}
  \label{eq:2-8}
  T^\Phi(x^\Phi) = \tilde T(\nabla\Phi(x) \nabla\Phi(x)^T)
\end{equation}
and
\begin{equation}
  \label{eq:2-9}
  \Sigma(x) = \tilde \Sigma(\nabla\Phi(x)^T \nabla\Phi(x)) \:.
\end{equation}
Note that both functions, $\tilde T$ and $\tilde \Sigma$ are invariant under
rigid motions. The first Piola-Kirchhoff stress, in contrast, can not be
expressed in a form depending only on $\nabla\Phi^T \nabla\Phi$. Nonetheless,
$T$ appears in the governing equations~(\ref{eq:2-5}) and so the forces $f,g$,
in contrast to $\tilde g$ appearing in~(\ref{eq:2-7}), are not invariant under
rotations. However, with regard to~(\ref{eq:2-6}) the difference is just a
factor of $\nabla\Phi^{-1}$. From now on we will not distinguish between the
stress tensors $T, \Sigma$ and their response functions $\tilde T,
\tilde\Sigma$ and we introduce the notations $F=\nabla\Phi$ and $C=F^TF$ for
the sake of brevity. Note that $C$, called Green strain, is the quantity that
describes the local change of distances of nearby points (a Taylor expansion
of $\norm{\Phi}{}^2$ can easily show this). Furthermore, one can show that two
deformations of a body that have the same Green strain differ from one another
by a rigid motion only. Hence, two deformations of the same body can be
identified if they have the same Green strain.

Suppose now, that the first Piola-Kirchhoff stress is the derivative of a
stored energy function $\hat W(F)$, i.e.,
\begin{equation}
  \label{eq:2-10}
  T(F) = \frac{\partial \hat W}{\partial F}(F) \:.
\end{equation}
A material with this property is called hyperelastic. Hyperelastic materials
are advantageous for two main reasons. First, if also the forces
in~(\ref{eq:2-5a}) are conservative, i.e., they can be written as the
G\^ateaux derivative of some potential, the equations of
equilibrium~(\ref{eq:2-5}) can be written as a minimization problem. Since the
forces will be unknown in our optimization problem this will be less important
to us. The second advantage is that they allow a much more intuitive modeling
of energetic penalties to certain kinds of deformations (e.g., non-isochoric
deformations) rather than a direct modeling via the response function.

The materials that we will use in the sequel are homogeneous and isotropic but
before formulating their stored energy functions recall the principal
invariants of a matrix $A\in\real^{3\times 3}$:
\begin{equation}
  \label{eq:2-11} 
  \begin{split}
    i_1(A) & = \tr{A} \:, \\
    i_2(A) & = \tr{\cof{A}} \:, \quad \text{and} \\
    i_3(A) & = \det{A} \:,
  \end{split}
\end{equation}
where $\tr{A}$ is the trace and $\cof{A} = (\det{A})A^{-T}$ is the cofactor
matrix (for invertible $A$). In terms of eigenvalues the principal invariants
can be written as
\begin{equation}
  \label{eq:2-12} 
  \begin{split}
    i_1(A) & = \lambda_1+\lambda_2+\lambda_3 \:, \\
    i_2(A) & = \lambda_1\lambda_2+\lambda_2\lambda_3+\lambda_1\lambda_3 \:,
    \quad \text{and} \\
    i_3(A) & = \lambda_1\lambda_2\lambda_3 \:.
  \end{split}
\end{equation}
The so-called reduced principal invariants are given by
\begin{equation}
  \label{eq:2-13}
  \begin{split}
    I_1(A) & = i_3(A)^{-1/3}i_1(A) \:, \\
    I_2(A) & = i_3(A)^{-2/3}i_2(A) \:, \quad \text{and} \\
    J(A) & = i_3(A)^{1/2} \:.
  \end{split}
\end{equation}
For isotropic hyperelastic materials $\hat W$ can be formulated in terms of
the (reduced) principal invariants of the Green strain $C$ only. We will make
use of two hyperelastic materials in this work. The stored energy function of
Saint Venant-Kirchhoff materials (SVK) is given by
\begin{equation}
  \label{eq:2-14}
  \begin{split}
    \hat W_{\text{SVK}} & = \mu (i_1(C)-3) +
    \frac{\lambda+2\mu}{8}(i_1(C)-3)^2 \\
    & \qquad - \frac{\mu}{3}(i_2(C)-3) \\
    & = \frac{\lambda}{2}(\tr{E})^2  + \mu \tr{(E^2)}
  \end{split}
\end{equation}
where $2E=C-\id$, $C=F^TF$, and $\lambda, \mu\geq 0$ are the Lam\'{e}
constants. SVK materials are linear material models since the second
Piola-Kirchhoff stress is a linear function of $E$ ($C$ respectively):
\begin{equation}
  \label{eq:2-15}
  \Sigma(E) = \lambda (\tr E)\id + 2\mu E \:.
\end{equation}
A linearization of $E$ yields the well known Hooke law of linear elasticity
which is also called ``small deformation-small strain setting''. Their
potential function is given by
\begin{equation}
  \label{eq:2-14a}
  \hat W_{\text{Linear}} = \frac{\lambda}{2}(\tr{\eps})^2  + \mu \tr{(\eps^2)} \:,
\end{equation}
where $2\eps = (F+F^T)-2\id$. SVK materials, however, can be shown to be
first order approximations of isotropic materials in terms of $E$ and
therefore this model is called ``large deformation-small strain''.

Neo-Hookean materials have a stored energy function of the form
\begin{equation}
  \label{eq:2-16}
  \hat W_{\text{NEO}} = \alpha (I_1(C)-3) + \beta (J(C)-1)^2
\end{equation}
where $\alpha,\beta>0$ are positive constants and are used to model rubber and
elastomers~\cite{Dhondt}. Sometimes different variants of the term involving
$J$ are described in the literature. The use of the reduced invariants ensures
the positivity of the stored energy. This model is the simplest model that is
appropriate for large strain and large deformations. Using~(\ref{eq:2-10})
equation~(\ref{eq:2-5}) can be rewritten for a general hyperelastic material
as
\begin{equation}
  \label{eq:2-17}
  \begin{split}
    -\DIV \frac{\partial \hat W}{\partial F}(\nabla\Phi) & = f(x) \:, \quad x\in \mathcal{B} \:, \\
    \frac{\partial \hat W}{\partial F}(\nabla\Phi)n & = g(x) \:, \quad
    x\in \partial \mathcal{B} \:.
  \end{split}
\end{equation}

\subsection{Discretization Using FEM}\label{s-2-2}

Equation~(\ref{eq:2-17}) is essentially a system of nonlinear partial
differential equations (PDEs) with pure Neumann boundary conditions. Usually,
this system is additionally supplemented with Dirichlet boundary conditions
that prescribe the deformation of $\mathcal{B}$ on (parts of) its
boundary. For now, we will not take into account the Dirichlet boundary
conditions but later we will make a connection between the boundary forces and
the deformation on the boundary.

System~(\ref{eq:2-17}) is commonly discretized by means of the finite element
method (FEM)~\cite{Braess,Dhondt,LeTallec}. Applying the FEM to nonlinear
elasticity problems is in general not straightforward. For incompressible or
nearly incompressible problems, i.e., the volume change during the deformation
is zero or very small, mixed FEMs need to be employed. However, in this work
we will avoid this difficulty and allow volume changes. For the purpose of
comparing shapes differing by an unknown deformation (due to unknown forces)
this is a reasonable assumption and therefore it is possible to simply employ
piece-wise linear FEMs. We shall elaborate on this in the following.

FEMs rely on the variational form of the PDE, i.e., we multiply the first
equation in~(\ref{eq:2-17}) with a test function $\Psi$ in a space $V$ of
suitably smooth test functions and integrate the left-hand side by parts. Then
one looks for a solution $\Phi\in V$ such that
\begin{multline}
  \label{eq:2-18}
  \int_\mathcal{B} \frac{\partial \hat W}{\partial F}(\nabla\Phi):\nabla\Psi \drm x =
  \int_\mathcal{B} f\cdot\Psi \drm x \\ + \int_{\partial \mathcal{B}} g\cdot\Psi \drm S \quad \forall
  \: \Psi\in V \:.
\end{multline}
where $A:B:=\tr(A^TB)$ denotes the Frobenius scalar product. $V$ is usually
some Sobolev space that encodes the Dirichlet boundary conditions and
regularity requirements on $\nabla\Phi$. We will, however, not get into the
details of the analytical treatment of~(\ref{eq:2-18}) which can be found
in~\cite{Ciarlet,LeTallec}.

In order to find an approximate solution FEMs replace the function space $V$
by a finite dimensional subspace $V^h$. The function space $V^h$ includes an
approximation $\mathcal{B}^h$ of the domain $\mathcal{B}$. The problem is then
to find $\Phi^h\in V^h$ such that
\begin{multline}
  \label{eq:2-19}
  \int_{\mathcal{B}^h} \frac{\partial \hat W}{\partial F}(\nabla\Phi^h):\nabla\Psi^h \d
  x = \int_{\mathcal{B}^h} f\cdot\Psi^h \drm x \\ + \int_{\partial \mathcal{B}^h} g\cdot\Psi^h \drm S
  \quad \forall \: \Psi^h\in V^h \:.
\end{multline}
This problem is nonlinear in the deformation $\Phi^h$ and can be solved given
appropriate boundary conditions by Newton-Raphson type methods which require
linearization. Let $\Phi_0^h\in V^h$ be a given deformation. Then, the
linearized version of~(\ref{eq:2-19}) around $\Phi_0^h$ is given by
\begin{multline}
  \label{eq:2-20}
  \int_{\mathcal{B}^h} \frac{\partial \hat W}{\partial F}(\nabla\Phi_0^h):\nabla\Psi^h
  \drm x \\ + \int_{\mathcal{B}^h} \left(\frac{\partial^2 \hat W}{\partial
      F^2}(\nabla\Phi_0^h)(\nabla\Phi^h-\nabla\Phi_0^h)\right):\nabla\Psi^h \d
  x \\ = \int_{\mathcal{B}^h} f\cdot\Psi^h \drm x + \int_{\partial \mathcal{B}^h} g\cdot\Psi^h \drm S
  \quad \forall \: \Psi^h\in V^h \:.
\end{multline}
Note that the second derivative of $\hat W$ with respect to the deformation
gradient is a tensor of fourth order. A system of this type has to be solved
in each step of a Newton-Raphson procedure.

Next, in order to get to a more concrete representation, we choose a global
basis of the finite dimensional space $V^h$,
$(\Psi_i^l)_{i=1,\ldots,N}^{l=1,\ldots,3}$, so that testing~(\ref{eq:2-20})
with all functions of $V^h$ is equivalent to testing against all base
functions. We make the ansatz
\begin{equation}
  \label{eq:2-21}
  \Phi^h = \Phi_i^l \Psi_i^l \:, \quad i=1,\ldots,N \:, l=1,2,3 \:,
\end{equation}
with real coefficients $\Phi_i^l$ whereas we use Einstein's sum
convention. Note that the base functions are vector valued and $l$ stands for
the index of their relevant component. Let furthermore $\Phi_0^h
=\Phi_{0,i}^l\Psi_i^l$ be the expansion of $\Phi_0^h$ in the basis of
$V^h$. Plugging all this into~(\ref{eq:2-20}) we get
\begin{multline}
  \label{eq:2-22}
  \int_{\mathcal{B}^h} \frac{\partial \hat W}{\partial F}(\nabla\Phi_0^h):\nabla\Psi_j^m \drm x \\
  + \left(\Phi_i^l - \Phi_{0,i}^l\right)\int_{\mathcal{B}^h}
  \left(\frac{\partial^2 \hat W}{\partial
      F^2}(\nabla\Phi_0^h)\nabla\Psi_i^l\right):\nabla\Psi_j^m \drm x \\
  = \int_{\mathcal{B}^h} f\cdot\Psi_j^m \drm x + \int_{\partial
    \mathcal{B}^h} g\cdot\Psi_j^m \drm S \:.
\end{multline}
This is a linear equation for the nodal deformations $\Phi_i^l$ of the form
\begin{multline}
  \label{eq:2-23}
  b_{\Phi_0}(\Psi_j^m) + \Phi_i^l a_{\Phi_0}(\Psi_i^l,\Psi_j^m) = \hat f(\Psi_j^m) + \hat g(\Psi_j^m)
  \\ \forall j=1,\ldots,N \:, \quad m=1,2,3 \:.
\end{multline}
Here $b_{\Phi_0}$ corresponds to the term of zero order of the linearization
of~(\ref{eq:2-19}) and $a_{\Phi_0}$ is a bilinear form, corresponding to the
first order term of the linearization, whose representation matrix in terms of
the FEM base functions is called the tangent stiffness matrix. The right-hand
side represents a force term.

As mentioned above, we use piece-wise linear finite elements to approximate
$\Phi^h$. For this purpose the approximate geometry $\mathcal{B}^h$ is a
tetrahedralization of $\mathcal{B}$. A linear function in each tetrahedron
(tet) is then uniquely determined by its nodal values. A set of global base
functions is defined by the relation
\begin{equation}
  \label{eq:2-24}
  \Psi_j^l(x_k) = \delta_{jk}e^l \:, \quad j,k=1,\ldots,N \:,
\end{equation}
where $e^l$ is the $l$-th canonical base vector, $l=1,\ldots,3$, and $x_k$ are
the nodes of $\mathcal{B}^h$. Using this we can characterize $V^h$ as
\begin{multline}
  \label{eq:2-25}
  V^h = \left\lbrace \Psi \in C^0(\mathcal{B}^h,\real^3) \: | \: \text{$\Psi$ is linear} \right. \\
  \left. \text{in each tet of $\mathcal{B}^h$} \right\rbrace \:.
\end{multline}
This way each coefficient $\Phi_i^l$ in equation~(\ref{eq:2-21}) can be
interpreted as the $l$-th component of the deformation vector at the node
$x_i$ of the tessellation $\mathcal{B}^h$. Note, that the discretization
introduces two errors, the error due to the approximation of $\Phi\in V$, and
an error that is caused by the approximation of the geometry $\mathcal{B}$.

\section{Optimization of Boundary Conditions}\label{s-3}
	In this section we will give a detailed description of the optimization
procedure that we use to tackle the matching problem. We will show how to
approximate this nonlinear problem by a sequence of convex problems that can
be put in the form of a second order cone program (SOCP).

Our goal is to match two two-dimensional surface meshes, the source
$\mathcal{S}$ and the target $\mathcal{T}$, deviating from one another by an
unknown deformation $\Phi$. Very often these surfaces represent the surface of
actual physical entities like solid bodies. In order to deform solid bodies an
external force is necessary. As stated in the introduction, the external
forces we are seeking are simple, i.e., sparse and isotropic and act on the
boundary of the physical body $\mathcal{B}$ only. These forces are a priori
unknown as well as the deformation of the surface itself. But we can put them
in relation using the FEM, as we will show in the sequel. The optimization
problem of finding the deformation can be formulated as follows
\begin{equation}
  \label{eq:3-1}
  \begin{split}   
    \minimize_{\Phi} & \quad \int_{\partial\mathcal{B}} \norm{\frac{\partial\hat W}{\partial F}(\nabla\Phi)n}{2} \drm S \\
    \text{subject to} & \quad \DIV \frac{\partial \hat W}{\partial F}(\nabla\Phi) = 0 \:, \quad x\in \mathcal{B}  \\
      & \hspace{1.5cm} \Phi (\partial\mathcal{B}) = \mathcal{T} \:,
  \end{split}
\end{equation}
i.e., we are imposing elastic properties on the source
$\mathcal{S}=\partial\mathcal{B}$ and then we are seeking an elastic
deformation $\Phi$ of $\mathcal{B}$ such that the external forces that are
responsible for the deformation of $\mathcal{S}$ into $\mathcal{T}$ are
minimal. The objective in~(\ref{eq:3-1}) is essentially an $L^1$-cost on the
force. Note that we assume a pure boundary force and no volumetric forces. The
force term we optimize in the objective function of~(\ref{eq:3-1}) is measured
by the first Piola-Kirchhoff stress tensor, see equation~(\ref{eq:2-10}). They
are acting on the deformed configuration but are measured per unit area in the
reference configuration. Hence, the objective is not invariant under
rotation. This, however, is not a limitation because one can simply take the
result of the optimization~(\ref{eq:3-1}) and apply
transformation~(\ref{eq:2-6}) to get rotation invariant forces that are
measured and acting on the reference configuration. This way one can identify
two deformations that deviate from one another by a Euclidean transform.

Unfortunately, the optimization problem~(\ref{eq:3-1}) is nonlinear due to the
second constraint and due to the nonlinearity of material models. We will show
a way to approximate it by a sequence of simpler and, most importantly, convex
problems.

In contrast to the boundary forces one can often give a rough guess
$\Phi^\ast$ of surface correspondences between the surface meshes to be
compared. Again, regarding the source mesh $\mathcal{S}$ as the boundary of a
physical body this guess of correspondences induces a boundary force on
$\partial\mathcal{B}$. This boundary correspondence can then be optimized
according to our prior that the induced forces are sparse and isotropic. This
way we get a deformation of $\mathcal{S}$ that is caused by sparse forces and
is ``closer'' to the target and we can re-iterate the process. In each step we
have to solve a nonlinear optimization problem which can be formulated as
\begin{equation}
  \label{eq:3-2}
  \begin{split}   
    \minimize_{\Phi} & \quad \int_{\partial\mathcal{B}} \left(
      \norm{\frac{\partial\hat W}{\partial F}(\nabla\Phi)n}{2} \right. \\
    & \hspace{1.5cm} \left. + \frac{k}{2}\norm{\Phi-\Phi^\ast}{2}^2 \right) \drm S \\
    \text{subject to} & \quad \int_\mathcal{B} \frac{\partial \hat W}{\partial
      F}(\nabla\Phi):\nabla\Psi \drm x = \\ & \qquad \int_{\partial \mathcal{B}}
    \frac{\partial\hat W}{\partial F}(\nabla\Phi)n\cdot\Psi \drm S \\
    & \hspace{1.7cm} \forall
    \: \Psi\in V \:.
  \end{split}
\end{equation}
This way we eliminate the second constraint in~(\ref{eq:3-1}) by localizing
the optimization around the initial guess $\Phi^\ast$. The new term
in~(\ref{eq:3-1}) can be interpreted as a spring force with spring constant
$k$, measuring deviation from the estimated correspondences, that needs to be
balanced by sparse elastic forces. Note that we also replaced the PDE
constraint in~(\ref{eq:3-1}) by its variational form~(\ref{eq:2-18}).

Problem~(\ref{eq:3-2}) can be given a discrete formulation using
equation~(\ref{eq:2-19}), i.e.,
\begin{equation}
  \label{eq:3-3}
  \begin{split}
    \minimize_{\Phi} & \quad \int_{\partial\mathcal{B}^h} \left(
      \norm{\frac{\partial\hat W}{\partial F}(\nabla\Phi^h)n}{2} \right. \\
    & \hspace{1.5cm} \left. + \frac{k}{2}\norm{\Phi^h-\Phi^{h,\ast}}{2}^2 \right) \drm S \\
    \text{subject to} & \quad \int_{\mathcal{B}^h} \frac{\partial \hat
      W}{\partial F}(\nabla\Phi^h):\nabla\Psi^h \drm x = \\
    & \qquad \int_{\partial \mathcal{B}^h} \frac{\partial\hat
      W}{\partial F}(\nabla\Phi^h)n\cdot\Psi^h \drm S \\
    & \hspace{1.7cm} \forall \: \Psi^h\in
    V^h \:.
  \end{split}
\end{equation}
This is still a difficult nonlinear problem although now finite
dimensional. The constraint as an equation itself, i.e., the discretized
variational form of the nonlinear elasticity equations, is usually solved by
means of a Newton-Raphson type procedure that relies on successive
linearization of the nonlinear terms. This linearization suggests a way to
treat the boundary force term in the objective of~(\ref{eq:3-3}). In each
Newton-Raphson step we have to solve a linear system of the
type~(\ref{eq:2-23}). This system can be recast by a renumbering of the nodal
deformations into the form
\begin{equation}
  \label{eq:3-4}
  \begin{split}    
    \left[
      \begin{array}{c}
        \tilde G_B \\ 0 \\
      \end{array}
    \right]
    = &
    \left[
      \begin{array}{c}
        B_{B,\Phi_0^h} \\ B_{I,\Phi_0^h} \\
      \end{array}
    \right] \\
    & +
    \left[
      \begin{array}{cc}
        A_{BB,\Phi_0^h} & A_{BI,\Phi_0^h} \\
        A_{IB,\Phi_0^h} & A_{II,\Phi_0^h} \\
      \end{array}
    \right]
    \left[
      \begin{array}{c}
        \Phi_B \\ \Phi_I \\
      \end{array}
    \right] \:,
  \end{split}
\end{equation}
i.e., we decompose the deformation vector
\begin{equation}
  \label{eq:3-5}
  \Phi^h=( \Phi_1^1, \Phi_1^2, \Phi_1^3, \ldots, \Phi_N^1, \Phi_N^2, \Phi_N^3  )^T
\end{equation}
into a boundary deformation part $\Phi_B$ collecting all the deformation
vectors on boundary nodes of $\mathcal{B}^h$ and into a part $\Phi_I$ of
deformation vectors at the inner nodes. The right-hand side represents
external boundary forces $\tilde G_B$ and zero external volumetric forces
in~(\ref{eq:2-23}).

Equation~(\ref{eq:3-4}) can be used to replace the nonlinear constraint
in~(\ref{eq:3-3}):
\begin{equation}
  \label{eq:3-6}
  \begin{split}
    \minimize_{\Phi} & \int_{\partial\mathcal{B}^h} \norm{\frac{\partial\hat W}{\partial F}(\nabla\Phi^h)n}{2} \\ 
    & \qquad + \frac{k}{2}\norm{\Phi^h-\Phi^{h,\ast}}{2}^2 \drm S \\
    \text{subject to} & 
        \left[
      \begin{array}{c}
        \tilde G_B \\ 0 \\
      \end{array}
    \right]
    =  
    \left[
      \begin{array}{c}
        B_{B,\Phi_0^h} \\ B_{I,\Phi_0^h} \\
      \end{array}
    \right] \\
    & +
    \left[
      \begin{array}{cc}
        A_{BB,\Phi_0^h} & A_{BI,\Phi_0^h} \\
        A_{IB,\Phi_0^h} & A_{II,\Phi_0^h} \\
      \end{array}
    \right]
    \left[
      \begin{array}{c}
        \Phi_B \\ \Phi_I \\
      \end{array}
    \right] \:.
  \end{split}
\end{equation}
Note that the boundary force term $\tilde G_B$ in the linearized constraint is
essentially a smoothed version of the boundary force term
\begin{equation}
  \label{eq:3-7}
  \frac{\partial\hat W}{\partial F}(\nabla\Phi^h)n
\end{equation}
in the objective since it is weighted by the finite element base functions.

Equation~(\ref{eq:3-4}) is in a form such that it is possible to eliminate the
inner deformations $\Phi_I$ by taking the Schur complement with respect to the
block $A_{BB,\Phi_0^h}$, i.e.,
\begin{equation}
  \label{eq:3-8}
  \tilde G_B = B_{B,\Phi_0^h} - A_{BI,\Phi_0^h}A_{II,\Phi_0^h}^{-1}B_{I,\Phi_0^h} + S_{\Phi_0^h}\Phi_B
\end{equation}
where
\begin{equation}
  \label{eq:3-9}
  S_{\Phi_0^h} = A_{BB,\Phi_0^h} - A_{BI,\Phi_0^h} A_{II,\Phi_0^h}^{-1} A_{IB,\Phi_0^h} \:.
\end{equation}
Utilizing~(\ref{eq:3-8}) we can rewrite the optimization problem as
\begin{equation}
  \label{eq:3-10}
  \minimize_{\Phi} \sum_{i=1}^K \norm{\tilde G_{B,i}}{2} + \frac{k}{2}\norm{\Phi_B - \Phi_B^\ast}{2}^2
\end{equation}
which is a convex problem. Each $\tilde G_{B,i}$, $i=1,\ldots,K$, is a force
vector in $\real^3$ attached to the $i$-th of the $K$ boundary nodes. Note
that~(\ref{eq:3-10}) is unconstrained and that $G_{B,i}$ depends on $\Phi_0^h$
and linearly on $\Phi_B$. Also, the dimension of the (local) optimization
problem is significantly reduced since it only involves the boundary
deformations $\Phi_B$. Such condensation techniques have been used in case of
completely linear elasticity in~\cite{Bro-Nielsen,Simon}.

The derived local optimization problem~(\ref{eq:2-10}) describes a single step
in our optimization scheme and needs three initialization parameters:
\begin{itemize}
\item An initial deformation $\Phi_0^h:B^h\rightarrow\real^3$ needs to be
  given in order to assemble the tangent stiffness matrix. We set
  $\Phi_0^h=\id$ before the first step.

\item The spring constant $k$ determining the strength of the spring force. As
  the source gradually deforms into the target during the iterations the
  elastic force that resists the springs increases. Thus, one needs to
  increase $k$ as well.

\item An initial guess of correspondence
  $\Phi^{h,\ast}:\mathcal{S}\rightarrow\mathcal{T}$ determining the direction
  of the spring force. This initial guess becomes more accurate the ``closer''
  the source is to the target shape.
\end{itemize}

The last point suggests that it makes sense to constrain the search for
improved correspondences locally to the target shape $\mathcal{T}$. This is
done by increasing the spring force penalty for deviation from the estimated
correspondences into normal direction during the process and amounts to
replacing the Euclidean metric in the second term of~(\ref{eq:3-10}), i.e.,
\begin{equation}
  \label{eq:3-11}
  \norm{\Phi_B - \Phi_B^\ast}{2}^2 = \sum_{i=1}^K\norm{\Phi_B(x_i)-\Phi_B^\ast(x_i)}{2}^2
\end{equation}
with a Mahalanobis distance that depends on the corresponding point
$\Phi_B^\ast(x_i)\in\mathcal{T}$ where $x_i$ is a node of $\mathcal{S}$. Such
a local metric is described by a symmetric positive definite $3$-by-$3$ matrix
\begin{equation}
  \label{eq:3-12}
  Q_{\Phi_B^\ast(x_i)}
  =
  \left[
    \begin{array}{c}
      n^T \\ t_1^T \\ t_2^T 
    \end{array}
  \right]
  \left[
    \begin{array}{ccc}
      \lambda_n & 0 & 0 \\
      0 & \lambda_t & 0 \\
      0 & 0 & \lambda_t \\
    \end{array}
  \right]
  \left[
    \begin{array}{ccc}
      n & t_1 & t_2 \\
    \end{array}
  \right]
\end{equation}
where $n$ is the unit normal at $\Phi_B^\ast(x_i)\in\mathcal{T}$ and $t_1,
t_2$ form an orthonormal basis of the tangent space. We can then
write~(\ref{eq:3-11}) as
\begin{equation}
  \label{eq:3-13}
  \norm{\Phi_B - \Phi_B^\ast}{Q}^2 = \sum_{i=1}^K\norm{\Phi_B(x_i)-\Phi_B^\ast(x_i)}{Q_{\Phi_B^\ast(x_i)}}^2
\end{equation}
where
\begin{equation}
  \label{eq:3-14}
  \norm{x}{Q}^2 = \dual{x}{Qx}
\end{equation}
and $Q=\diag{Q_1,\ldots,Q_K}$. Decreasing the ratio $\lambda_n / \lambda_t$
then means increasing the penalty in normal direction.
Combining~(\ref{eq:3-13}) with~(\ref{eq:3-10}) and introducing slack variables
$g_i$ and $e$ we can reformulate~(\ref{eq:3-10}) as
\begin{equation}
  \label{eq:3-15}
  \begin{split}
    \minimize_{g_i,e,\Phi} & \sum_{i=1}^K g_i + \frac{k}{2} e \\
    \text{subject to} & \norm{\tilde G_{B,i}}{2}\leq g_i \;, i=1,\ldots,K \\
    & \norm{\Phi_B - \Phi_B^\ast}{Q}^2\leq e \:. \\
  \end{split}
\end{equation}
The first $K$ constraints are already in second order cone form and the
constraint on $e$ can be reformulated using
\begin{equation}
  \label{eq:3-16}
  \norm{\Phi_B - \Phi_B^\ast}{Q}^2\leq \frac{1}{4}\left( (e+1)^2 - (e-1)^2  \right)
\end{equation}
into
\begin{equation}
  \label{eq:3-17}
  \norm{
    \left[
      \begin{array}{c}
       2R(\Phi_B - \Phi_B^\ast) \\ e-1
       \end{array}
     \right]
   }{2}
   \leq
   e
\end{equation}
where $R = \diag{Q_1,\ldots,Q_K}^{1/2}$. The SOCP in each iteration then
becomes
\begin{equation}
  \label{eq:3-18}
  \begin{split}
    \minimize_{g_i,e,\Phi} & \sum_{i=1}^K g_i + \frac{k}{2} e \\
    \text{subject to} & \norm{\tilde G_{B,i}}{2}\leq g_i \;, i=1,\ldots,K \\
    & \norm{ \left[
        \begin{array}{c}
          2R(\Phi_B - \Phi_B^\ast) \\ e-1
        \end{array}
      \right] }{2} \leq e \:.
  \end{split}
\end{equation}
There are several efficient solvers for SOCPs. We found that
MOSEK~\cite{MOSEK} is very suitable and used it through the YALMIP
interface~\cite{YALMIP} with MATLAB.

\section{Implementation and Results}\label{s-4}
	\subsection{The Two-scale Algorithm}\label{s-4-1}

In this section we describe our algorithm and its implementation aspects in
detail. As input our method takes two surface meshes
$\mathcal{S},\mathcal{T}$. The source $\mathcal{S}$ is assumed to represent
the surface of a deformable body. Elastic properties on $\mathcal{S}$ can then
be imposed on a volumetric tetrahedralization $\mathcal{B}^h$. Our formulation
so far assumes that the boundary of the tetrahedralization coincides with the
surface mesh. Tetrahedral meshes that respect the given surface tessellation
$\mathcal{S}$ can be obtained, for example, with TetGen~\cite{TetGen}. This,
however, has two disadvantages. First, the mesh $\mathcal{S}$ is required to
be free of artifacts like self-intersecting triangles or non-manifold
edges. This is, in general, not the case for meshes obtained from scanned
data. Secondly, a tet mesh with the same boundary as $\mathcal{S}$ will have a
large number of degrees of freedom that are relevant in our model and although
the optimization procedure is effectively reduced to the boundary of
$\mathcal{B}^h$ this is computationally very expensive. We can remedy this by
measuring elastic properties on a coarser scale than the spring force
in~(\ref{eq:3-10}), as we explain below.

To this end we created a coarse tetrahedralization $\mathcal{B}^h$, typically
with 2000 to 3000 tets (compared to the surface meshes that contained between
10000 to 50000 triangles). We used the Iso2Mesh toolbox~\cite{Iso2Mesh} that
wraps code provided in~\cite{CGAL} and~\cite{TetGen}. Next, we extracted the
boundary mesh of $\mathcal{B}^h$ and projected every node in $\mathcal{S}$
onto the mesh $\partial\mathcal{B}^h$, i.e., for each node $p\in\mathcal{S}$
we computed the triangle $t \in \partial\mathcal{B}^h$ with minumum distance
and the corresponding projection point $x_t \in t$. We then computed its
barycentric coordinates $(u,v,w)$ in $t$. The deformation of $p$, $\Phi(p)$,
can then be written as a linear combination
\begin{equation}
  \label{eq:4-1}
  \Phi(p) = u\vphi_B^h (x_1) + v\vphi_B^h (x_2) + w\vphi_B^h (x_3)
\end{equation}
of the deformation of the nodes $x_1,x_2,x_3$ of $t$. From now on we will
therefore distinguish between the deformation
$\Phi:\mathcal{S}\rightarrow\real^3$ and the deformation
$\vphi_B^h:\partial\mathcal{B}^h\rightarrow\real^3$ of the boundary of the tet
mesh. An equation of the form~(\ref{eq:4-1}) applies to each displacement
vector $\Phi(p)$ for any $p\in\mathcal{S}$. Therefore, any deformation of
$\partial\mathcal{B}^h$ can be mapped linearly to a deformation of
$\mathcal{S}$. \myalgorithm

In addition we smooth the interpolated deformation field with a number of
damped Jacobi relaxations using the Laplace-Beltrami operator
$\Delta_{\mathcal{S}}$. Note that smoothing the (linearly interpolated)
deformation field on $\mathcal{S}$ amounts to smoothing the mesh
$\Phi(\mathcal{S})$. Interpolation and relaxation can be represented by a
linear operator $T$ that acts on the space of continuous functions
$C^0(\partial\mathcal{B}^h,\real^3)$ and maps into $C^0(\mathcal{S},\real^3)$,
i.e., we can write
\begin{equation}
  \label{eq:4-2}
  \Phi = T\vphi_B^h \:.
\end{equation}

We note that an alternative approach to extending a deformation from a coarse
to a fine scale was taken by Kovalsky et~al.~\cite{Kovalsky} and
in~\cite{Sumner}. There the map $T$ was determined by a moving least squares
optimization, i.e., each deformation vector $\Phi(p)$ on $\mathcal{S}$ is the
weighted average of nearby deformation vectors on $\mathcal{B}^h$. This way
local volumetric deformation of $\mathcal{B}^h$ induces a local deformation on
$\mathcal{S}$ whereas we use a surface deformation to surface deformation
approach.

We can now reformulate problem~(\ref{eq:3-18}) as follows,
\begin{equation}
  \label{eq:4-3}
   \begin{split}
     \minimize_{g_i,e,\vphi_B^h,\Phi} & \sum_{i=1}^K g_i + \frac{k}{2} e \\
     \text{subject to} & \norm{\tilde G_{B,i}}{2}\leq g_i \;, i=1,\ldots,K \\
     & \norm{ \left[
         \begin{array}{c}
           2R(\Phi - \Phi^\ast) \\ e-1
         \end{array}
       \right] }{2} \leq e \\
     & \Phi = T\vphi_B^h
  \end{split}
\end{equation}
where
\begin{equation}
  \label{eq:4-4}
  \begin{split}
    \tilde G_B = & B_{B,\vphi_0^h} -
    A_{BI,\vphi_0^h}A_{II,\vphi_0^h}^{-1}B_{I,\vphi_0^h} +
    S_{\vphi_0^h}\vphi_B^h
  \end{split}
\end{equation}
in the spirit of~(\ref{eq:3-8}). This is an optimization problem on two scales
since it measures the (computationally expensive) deformation cost for the
volumetric body on a coarse scale while the spring force between source and
target shape is measured on a fine scale. The algorithm is summarized in
Algorithm~\ref{alg1}.

\begin{figure*}[t!]
  \centering
  \includegraphics[width=0.95\textwidth]{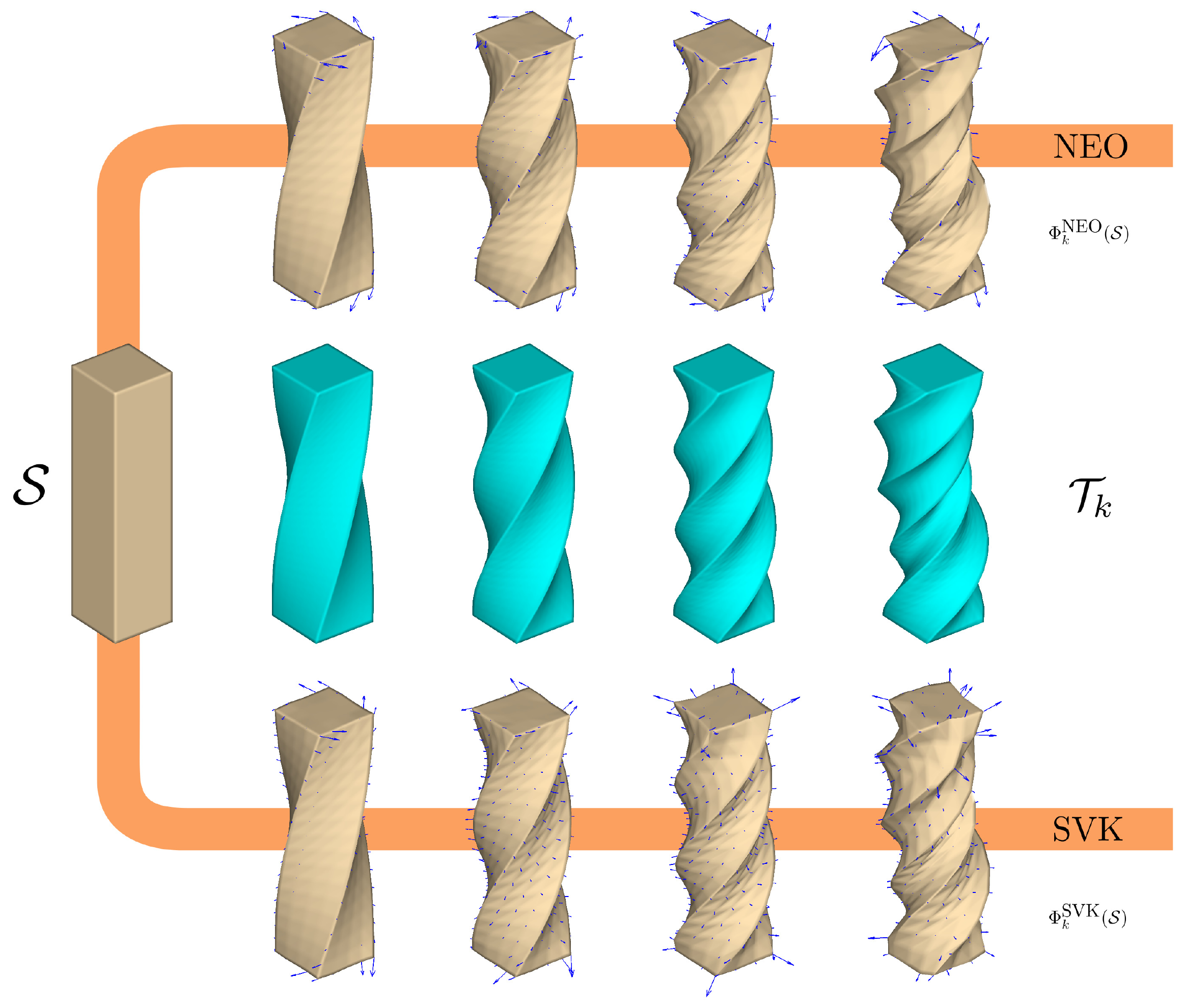}
  \caption{Deformation tracking experiment. The deformation sequence consists
    of 60 frames $\mathcal{T}_k$. In each frame the upper face of the beam
    $\mathcal{S}$ was twisted until a 360 degree twist was reached. The source
    $\mathcal{S}$ was then matched to each frame $\mathcal{T}_k$ while storing
    the result $\Phi_k(\mathcal{S})$ and gradually increasing the
    deformation. We show the result of the matching for four frames
    (90$^\circ$, 180$^\circ$, 270$^\circ$ and 360$^\circ$ twist) for the
    Neo-Hookean material model (upper row) and the Saint Venant-Kirchhoff
    Model (lower row). The optimized force vectors, measured on the coarse
    mesh $\mathcal{B}^h$, are shown on the results. The middle row shows the
    corresponding target frames (turquoise). The surface mesh $\mathcal{S}$
    consisted of 14400 triangles while the tetrahedral mesh $\mathcal{B}^h$ on
    which the forces of the deformation are measured consisted of 1536 tets
    (567 surface triangles on $\partial\mathcal{B}^h$).}
  \label{fig-1}
\end{figure*}

Our Matlab implementation was run on a standard desktop PC with 8GB of RAM and
Intel Core i7 with 3GHz. The bottleneck routines such as the assembly of the
tangent stiffness matrix were written in C. For SVK materials we wrote our own
FEM code and for Neo-Hookean materials we modified the source code of
CalculiX~\cite{CalculiX}, a three-dimensional FEM software that is available
under the GNU General Public License.

\vspace{0.5cm}

\textbf{Computing the Initial Guess.} Computing a reasonable initial guess is
not straightforward and is mostly done in a heuristic manner. In the case that
$\mathcal{S}$ and $\mathcal{T}$ only deviate by small deformations and
rotations it is often reasonable to find initial correspondences simply by
projecting points of the source mesh onto the target as done in a standard
iterative closest point algorithm. In the case of large deformations and/or
large rotations, however, nearest neighbor projection is usually a bad
choice. The limitation of small deformations and rotations can partially be
overcome by the use of shape descriptor like the heat kernel signature
(HKS)~\cite{HKS} or wave kernel signature (WKS)~\cite{WKS}. These descriptors
describe the diffusion of heat at each point or the dispersion in case of the
WKS. Since these descriptors are based on the spectrum of the Laplace-Beltrami
operator they are invariant under isometries, i.e., surface deformations that
preserve geodesic distances. For large isometric or nearly isometric
deformations one can then use a nearest neighbor search in the descriptor
space for the first few iterations of Algorithm~\ref{alg1} instead of the
Euclidean nearest neighbors. However, the so obtained correspondences can be
treacherous, in particular in the presence of inner symmetries of the shape.

In our work we use a descriptor guided nearest neighbour search in the
Euclidean space. As a shape descriptor we used the HKS. In the first few
iterations we seek for each point on the source shape $\mathcal{S}$ k-nearest
neighbours and then we select the one with the best matching HKS. We further
assign a confidence value $w\in [0,1]$ for the correspondence and use only
correspondences with confidence above a certain threshold. Similar weighting
procedures have been used in~\cite{Kovalsky,Choi,Rusinkiewicz}. The number $k$
of computed nearest neighbors was usually set around two to four percent of
the number of nodes of the shape. After a few iterations we decreased $k$ down
to one and passed on the confidence weighting what amounts to standard
Euclidean nearest neighbours.

\subsection{Experiments}\label{s-4-2}

\textbf{Material Models.} For our experiments we used the three material
models introduced in Section~\ref{s-2}. In the following we will refer to the
small deformation model~(\ref{eq:2-14a}) as the \emph{linear} model, to the
Saint Venant-Kirchhoff model~(\ref{eq:2-14}) as \emph{SVK} and to the
Neo-Hookean model~(\ref{eq:2-16}) as \emph{NEO}.

The first experiment is shown in Figure~\ref{fig-1} and is supposed to
demonstrate the flexibility of the two large deformation models (SVK and
NEO). We took a triangular surface mesh $\mathcal{S}$ of a beam and created a
sequence of deformations $\mathcal{T}_k$ (which we will call frames). In each
frame we twisted the upper face of the beam while keeping the bottom face
fixed until a twist of 360 degree was reached. We then used the frames
$\mathcal{T}_k$ as target frames for matching $\mathcal{S}$ according to
Algorithm~\ref{alg1}. We start from frame $k=1$ and gradually increase the
deformation while storing the resulting deformations $\Phi_k(\mathcal{S})$
until we reach the last frame. This is a very large deformation that causes
high mechanical stress. Note that all mechanical quantities are measured with
respect to the undeformed configuration $\mathcal{S}$.
\begin{figure}[h!]
  \centering
  \includegraphics[width=0.5\textwidth]{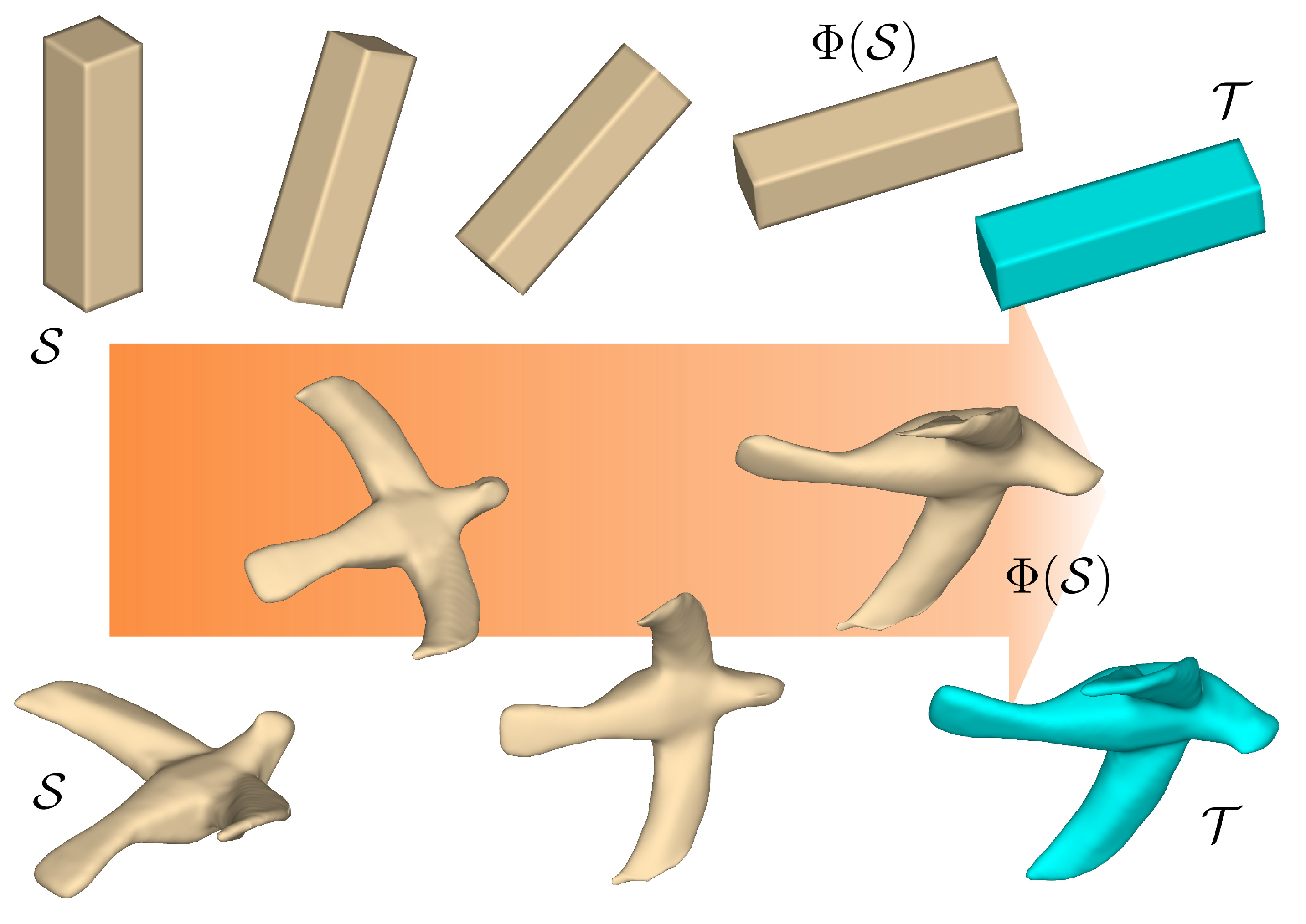}
  \caption{Our algorithm can find rigid motions. The target meshes
    $\mathcal{T}$ (transparent green, right) are rotations of the source
    meshes $\mathcal{S}$ (left). The bird mesh was taken from the SHREC data
    set. Our tet mesh had approximately 1400 tets compared to 18000 triangles
    of the surface mesh.}
  \label{fig-2}
\end{figure}

\begin{figure}[h!]
  \centering
  \includegraphics[width=0.5\textwidth]{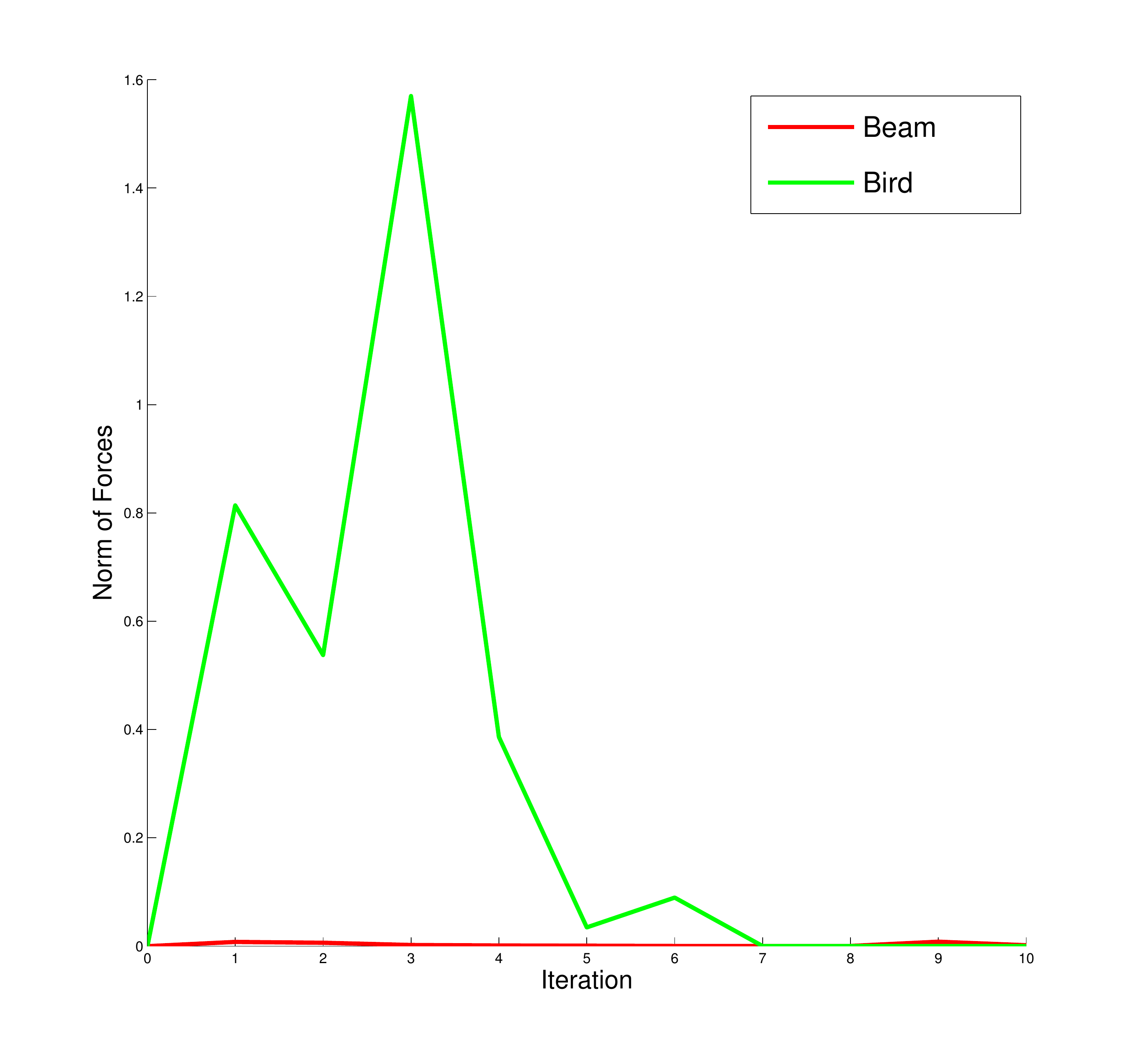}
  \caption{Discrete $L^1$-norm of the boundary forces in each
    iteration. The search space for the deformations is not confined
    to rigid motions but the deformation is ``not far'' from a rigid
    motion.}
  \label{fig-2g}
\end{figure}
Here, we did not perform the smoothing described in step~3 in
Algorithm\ref{alg1}. In the first few steps both materials performed
well. After a twist of approximately 270 degree we observed that the SVK model
needed more iterations in each step to converge than the Neo-Hookean
model. Also, the Newton-Raphson solver that is applied after every single
optimization step needed more iterations to converge to an actual solution of
the elasticity equations. The Neo-Hookean model, in contrast, showed a
substantial computational robustness against this large deformation. The
linear material model, i.e., the SVK model with linearized strain (no update
of the tangent stiffness matrix in step~11 of Algorithm~\ref{alg1}) failed to
produce reasonable results much earlier. However, we do not show the linear
case here since this material model is not suitable for this large deformation
anyway.

\begin{figure*}[h!]
  \centering
  \includegraphics[width=0.95\textwidth]{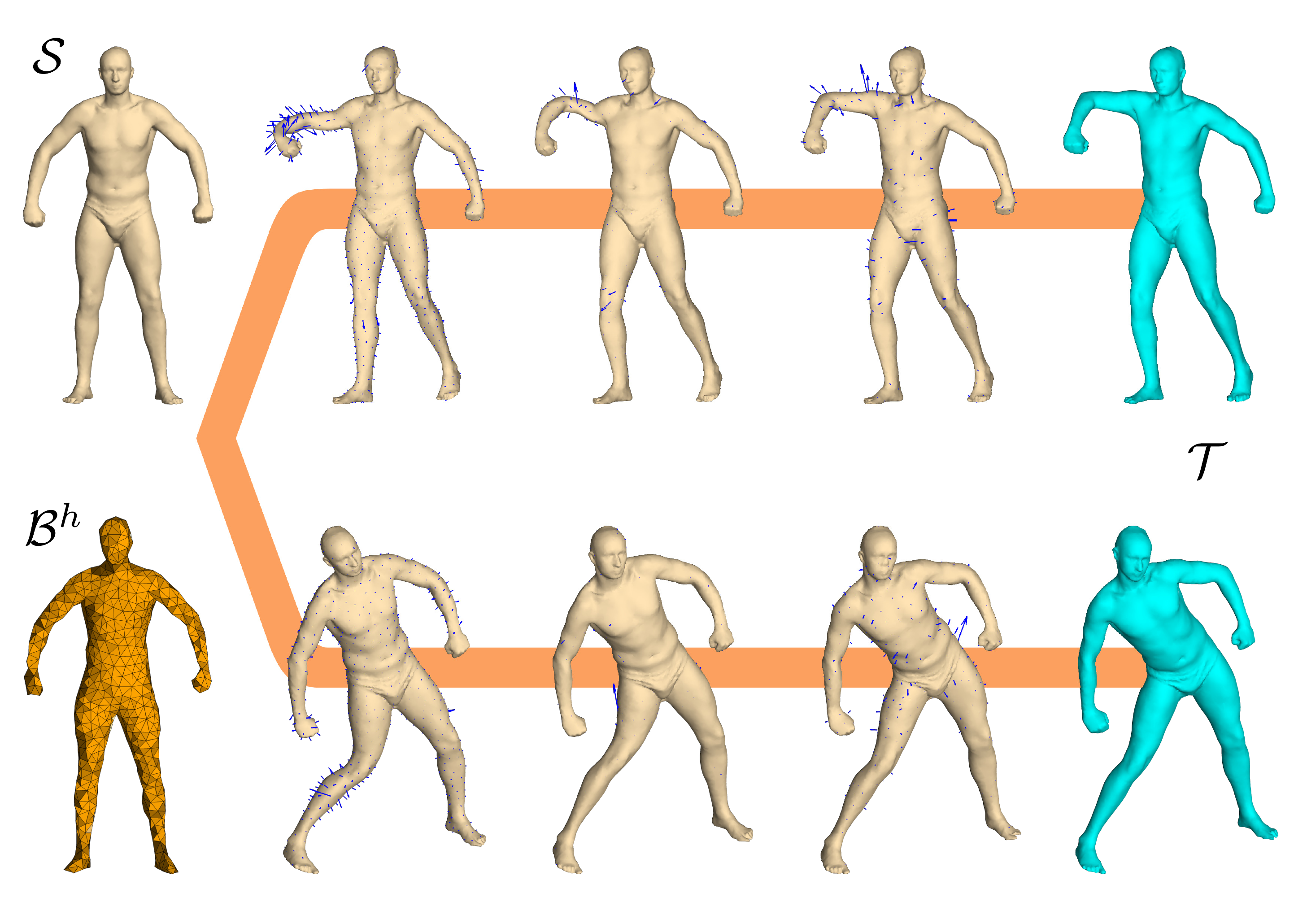}
  \caption{Several steps of the matching of a source shape (left, gray) to two
    different articulated human shapes (right, green) from the SCAPE
    dataset~\cite{SCAPE} are shown in the lower and upper row. The
    approximating volumetric mesh is shown on the left (orange) as well as the
    force vectors measured on the volumetric mesh. The force vectors are
    scaled for better visibility. The surfaces have approximately 25000
    triangles. Our volumetric tessellation had approximately 2000 tets.}
  \label{fig-3}
\end{figure*}

This coincides with our expectation since the SVK model is appropriate for
large deformations but not suitable for large strain in contrast to the
Neo-Hookean model which is suitable for both.

\vspace{0.5cm}

\textbf{Invariances.} As elaborated in Section~\ref{s-2} for each nonlinear
material the deformation force measured by the second Piola-Kirchhoff
stress~(\ref{eq:2-6}) is invariant under rotations. However, in our numerical
scheme we optimize a smoothed version of forces derived from the first
Piola-Kirchhoff stress which is, in general, not invariant under
rotations. Nevertheless, the magnitude of the forces is invariant but not the
direction. For pure rotations the force is zero. Hence, we would like the
matching algorithm to find pure rotations and to return a zero
force. Figure~\ref{fig-2} shows two examples. The initial guess of
correspondences for the bird was made and updated using a nearest neighbour
search in descriptor space in the first few iterations of our algorithm. For
the rotated beam we simply used Euclidean projections since the HKS is ``less
descriptive'' due to the high number of symmetries and hence might lead to an
unreasonable initial guess.

The graphs in Figure~\ref{fig-2g} show the norm of the optimized boundary
force in each iteration of the optimization. One can clearly see that in the
first few iterations the optimized boundary force is not zero. This is to be
expected since we do not restrict the search space of deformations to
rotations. In the last few iterations the force is close to zero implying that
our algorithm found a rigid motion of $\mathcal{B}^h$. The slightly incorrect
matching of the rotated surface mesh of the bird, as seen in
Figure~\ref{fig-2} (bottom right), is due to the interpolation of the
deformation of $\mathcal{B}^h$ to $\mathcal{S}$ and can be remedied by taking
a finer tet mesh at the cost of computation time.

Last, we should mention that in case of a linear material model (Hooke's law)
the forces are not invariant under rotations but under infinitesimal
rotations. These maps are the generators of the rotation group, i.e., they
form the Lie algebra of $SO(3)$, and have skew symmetric derivatives. In two
dimensions the effects of this invariance were demonstrated
in~\cite{Simon}. Here, we will refrain from showing this.

\begin{figure*}[h!]
  \centering
  \includegraphics[width=0.95\textwidth]{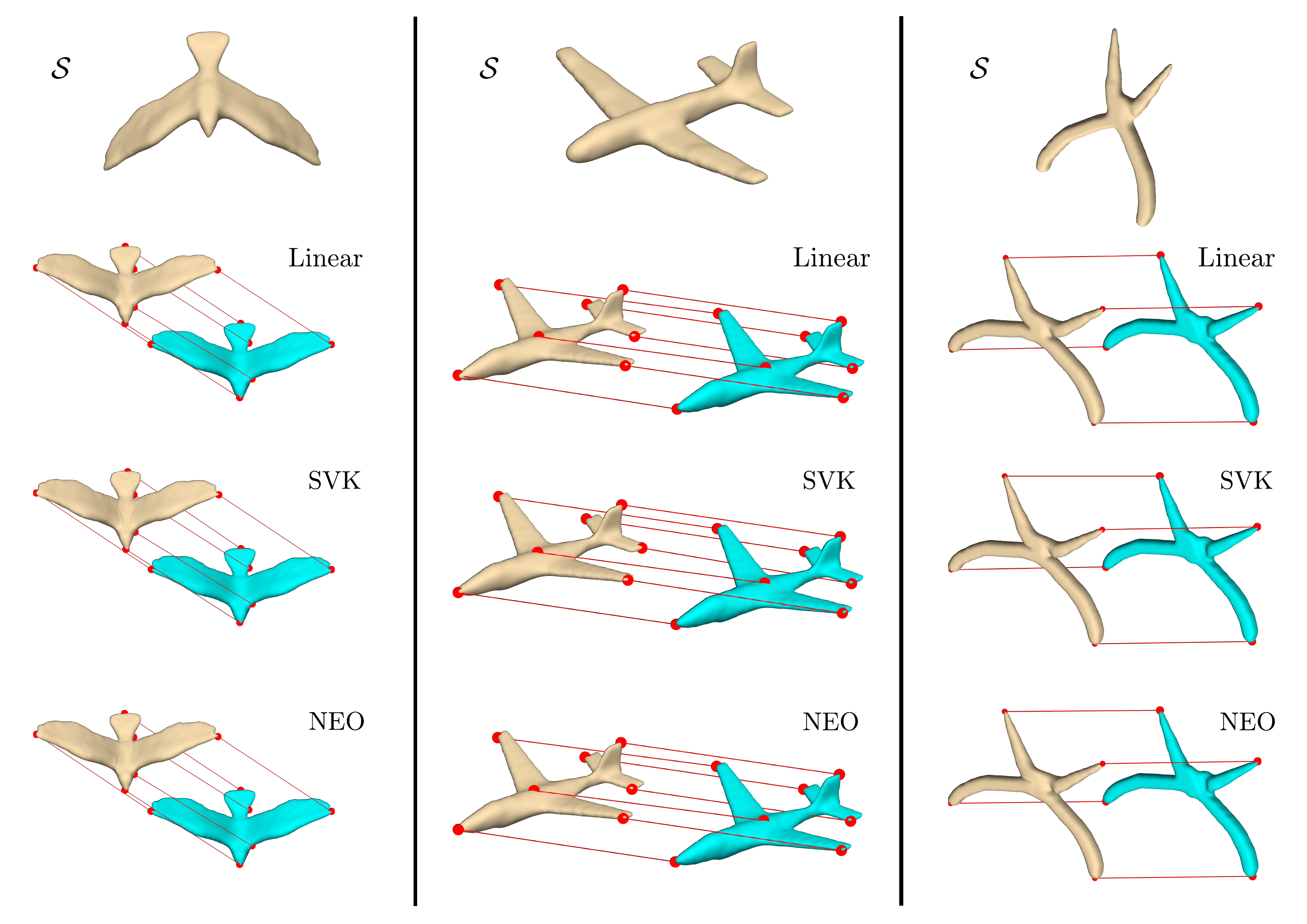}
  \caption{Examples from the SHREC dataset~\cite{SHREC}. The source shapes are
    shown in the top row. We compare three material models: the linear model,
    the Saint Venant-Kirchhoff model (SVK) and the Neo-Hookean model
    (Neo). Several manually labeled landmark points of the source meshes are
    shown on the deformed source mesh and then compared to their corresponding
    landmarks in the target shapes (green). The SHREC surface meshes we used
    consist of 9000 to 14000 triangles and the tetrahedral meshes of 1600 to
    2700 tets.}
  \label{fig-5}
\end{figure*}

\vspace{0.5cm}

\textbf{More Experiments.} Figure~\ref{fig-3} shows two examples of matching
human poses. The models were taken from the SCAPE dataset~\cite{SCAPE}. Both
poses were aligned using the SVK model. Note that the SCAPE dataset is nearly
isometric and hence the HKS is nearly invariant under change of poses. This
facilitates the nearest neighbor search in the first few iterations of our
algorithm. Note that the external force found by our algorithm is sparse and
mostly acts on the articulated parts of the model.

Examples of matching taken from the SCHREC dataset~\cite{SHREC} are shown in
Figure~\ref{fig-5}. Note that this dataset consists of usually non-isometric
instances within each object class. Still, our algorithm produces good
results.
\begin{table}[h]
  \centering
  \begin{tabular}{ c | c | c | c }
    & Linear & SVK & NEO \\ \hline
    \textbf{Birds}&  & &  \\ \hline
    Norm of forces & 6  & 14 & 2 \\ \hline
    \# Iterations & 14 & 35 & 36 \\
    \hline \hline
    \textbf{Planes} &  &  &  \\ \hline
    Norm of forces & 6.01  & 9.4 & 7.9 \\ \hline
    \# Iterations & 17 & 21 & 36 \\
    \hline \hline
    \textbf{Pliers} &  &  &  \\ \hline
    Norm of forces & 0.6  & 4.7 & 0.8 \\ \hline
    \# Iterations & 7 & 7 & 11 \\
    \hline
  \end{tabular}
  \caption{Comparison of the three material models used on the examples of
    Figure~\ref{fig-5}. We show the sum of the norms of the force vectors
    and the number of iterations}
  \label{tab:1}
\end{table}

\begin{figure*}[h!]
  \centering
  \includegraphics[width=0.95\textwidth]{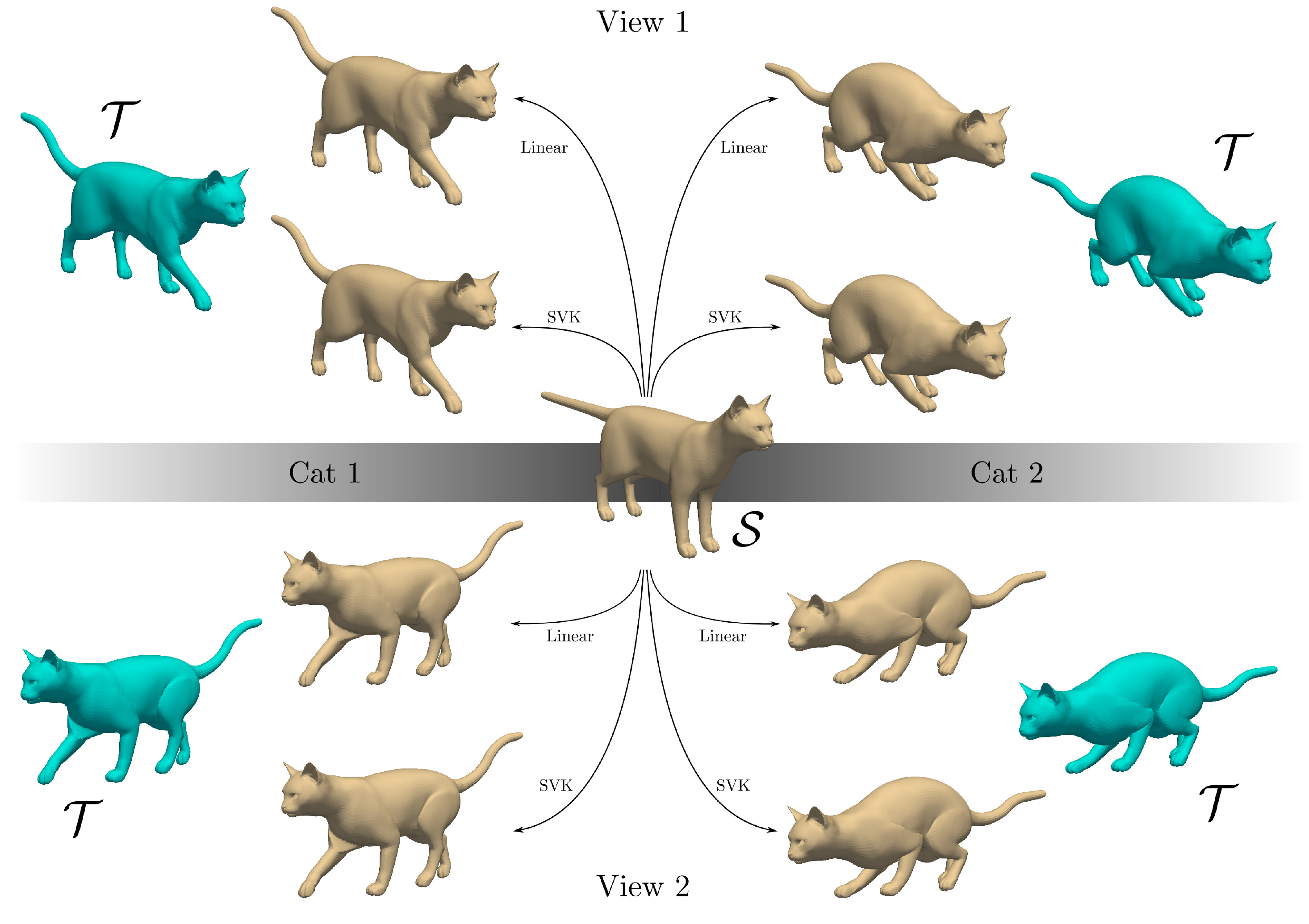}
  \caption{Two examples from the TOSCA dataset~\cite{TOSCA} from two different
    viewpoints (upper and lower row). The source (middle) is matched to the
    targets (green). Each shape in the TOSCA dataset consists of approximately
    55000 triangles. The volumetric mesh we used consists of 1500 tets. Both
    results are of similar quality whereas the SVK model needed forces of
    higher magnitude.}
  \label{fig-4}
\end{figure*}

Two examples taken from the TOSCA dataset~\cite{TOSCA} (left and right poses)
are shown in Figure~\ref{fig-4} from two different viewpoints (top and
bottom). We used the SVK and the linear model, i.e., the SVK model without
update of the tangent stiffness matrix (Hooke's law), for each pose. Both
results exhibit similar quality whereas the SVK model exhibits deformation
forces that are higher in magnitude. This is in line with the comparison of
the SVK model and the linear model in the experiment shown in
Figure~\ref{fig-5}.

\begin{figure*}[t!]
  \centering
  \includegraphics[width=0.95\textwidth]{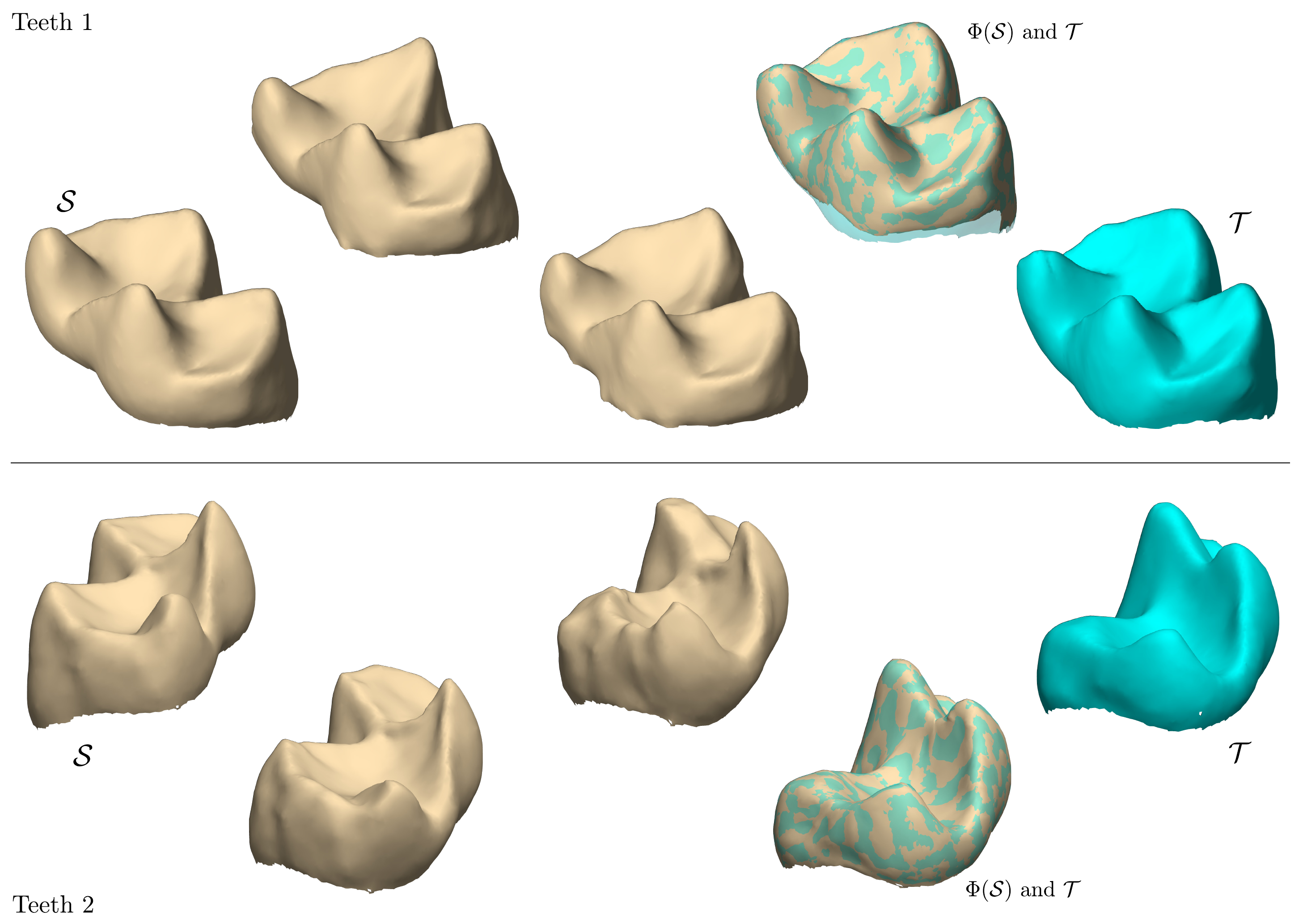}
  \caption{Two experiments performed on tooth shapes taken from the dataset of
    anatomical shapes of~\cite{ANATOMY} (top and bottom row). The meshes have
    disk topology and the deformations are nonisometric and exhibit large
    strain. Several intermediate steps of the matching procedure of the
    sources (left, gray) into the target (right, green) are shown. We used the
    Neo-Hookean model. Each mesh we used has about 10000 triangles. The
    volumentric meshes that we created had about 1000 tets.}
  \label{fig-6}
\end{figure*}

The experiments performed so far have been carried out on closed surfaces
without boundary. However, the method is not restricted to closed surfaces
provided one can create a volumetric mesh that represents the given object. To
create the volumetric mesh it is usually necessary to fill holes. This,
however, is the only step at which we use the closed surface.  The rest of the
steps follows the steps described in Algorithm~\ref{alg1}. We demonstrate this
in Figure~\ref{fig-6}. There we took two models of teeth from the anatomical
dataset of~\cite{ANATOMY} which exhibits quite large deformations. The
deformation of the underlying tetrahedral mesh $\mathcal{B}^h$ is shown in
Figure~\ref{fig-7}.

\section{Discussion}\label{s-5}
	In this work we propose a new ICP-like method in combination with a nonlinear
regularizer for three-dimensional surface matching. The idea is to regard the
surface as the surface of a physical body that can undergo elastic
deformation. Our approach to finding a reasonable deformation is led by the
assumption that the forces acting on the body are of a simple nature, i.e., we
assume the forces are sparse, isotropic and act on the boundary only. This is,
we believe, a reasonable assumption in many scenarios of shape deformation
such as the change of a human pose.

We model the physical body as a hyperelastic material. The underlying
equations, a nonlinearly coupled system of PDEs, is solved by means of a
conformal FEM and needs to be supplemented with appropriate boundary
conditions. We provide pure Dirichlet boundary conditions, i.e., the boundary
deformation. This boundary deformation is in general unknown but one can give
an initial guess. This initial guess induces a spring force between the source
and the target that is supposed to drive the deformation of the source. We
then use our proposed method to improve the initial guess such that the forces
that act on the deformed body are sparse and isotropic according to the
prior. This is done using the FEM which allows us to split the tangent
stiffness matrix. The splitting we employ allows to connect boundary forces on
the deformed body and boundary deformation of the source. This splitting
generalizes the linear condensation methods that have been used
in~\cite{Bro-Nielsen,Simon} to the nonlinear case.

The optimization problem to be solved is nonlinear and computationally
complex, depending on the shape and on the material model. We show how to
approximate this problem by a sequence of convex problems that is solved in an
iterative fashion with low computational demands. The convexification is done
by successive linearization of the material model. The computational
complexity is lowered by approximating the source surface with a coarser
tetrahedral mesh. A deformation of the boundary of the volumetric tetrahedral
mesh then naturally induces a deformation of the source surface by means of
interpolation and smoothing. This amounts to a two-scale algorithm since the
elastic properties are measured on the coarse scale while the provided initial
guess
\begin{figure}[h!]
  \centering
  \includegraphics[width=0.5\textwidth]{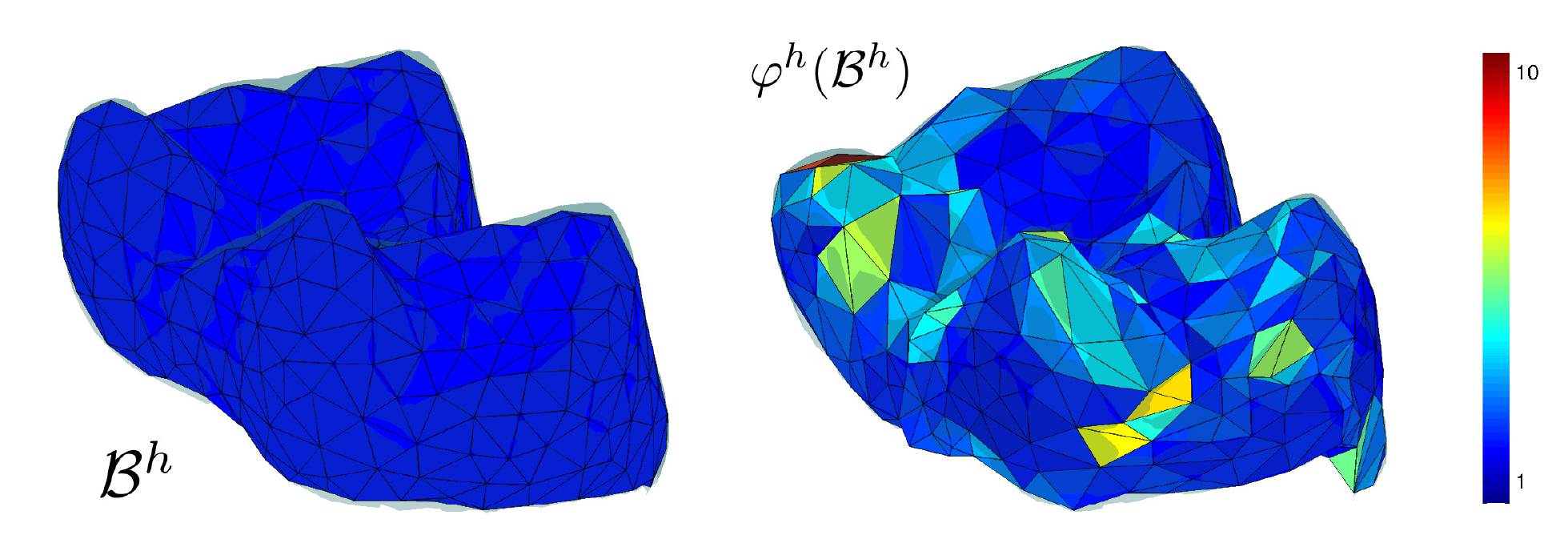}
  \caption{The volumetric mesh in rest state approximating the source shape
    shown in the top row of Figure~\ref{fig-6} is shown on the left. The right
    side shows the deformed mesh after matching including the conformal
    distortion of each tet.}
  \label{fig-7}
\end{figure}
is a map on the (fine) source surface. Reducing the optimization problem to
the boundary of the coarse scale volumetric mesh further reduces the
dimensionality of the problem significantly.

Our method often delivers good results but can fail in certain cases. The
estimated correspondences that need to be given before each iteration of the
optimization is, at least in the first few iterations of our algorithm,
usually computed by a Euclidean $k$-nearest neighbor search. Out of these $k$
nearest neighbors we pick the one with best matching HKS. In case of strongly
non-isometric deformations or shapes with a high amount of intrinsic symmetry
this will result in very unreasonable initial correspondences and the
algorithm will get stuck in a local minimum. The case $k=1$ is the usual
Euclidean nearest neighbor search and will give unreasonable correspondences
in case of large deformations and/or rotations even if the deformation is
isometric.

\begin{figure}[h!]
  \centering
  \includegraphics[width=0.5\textwidth]{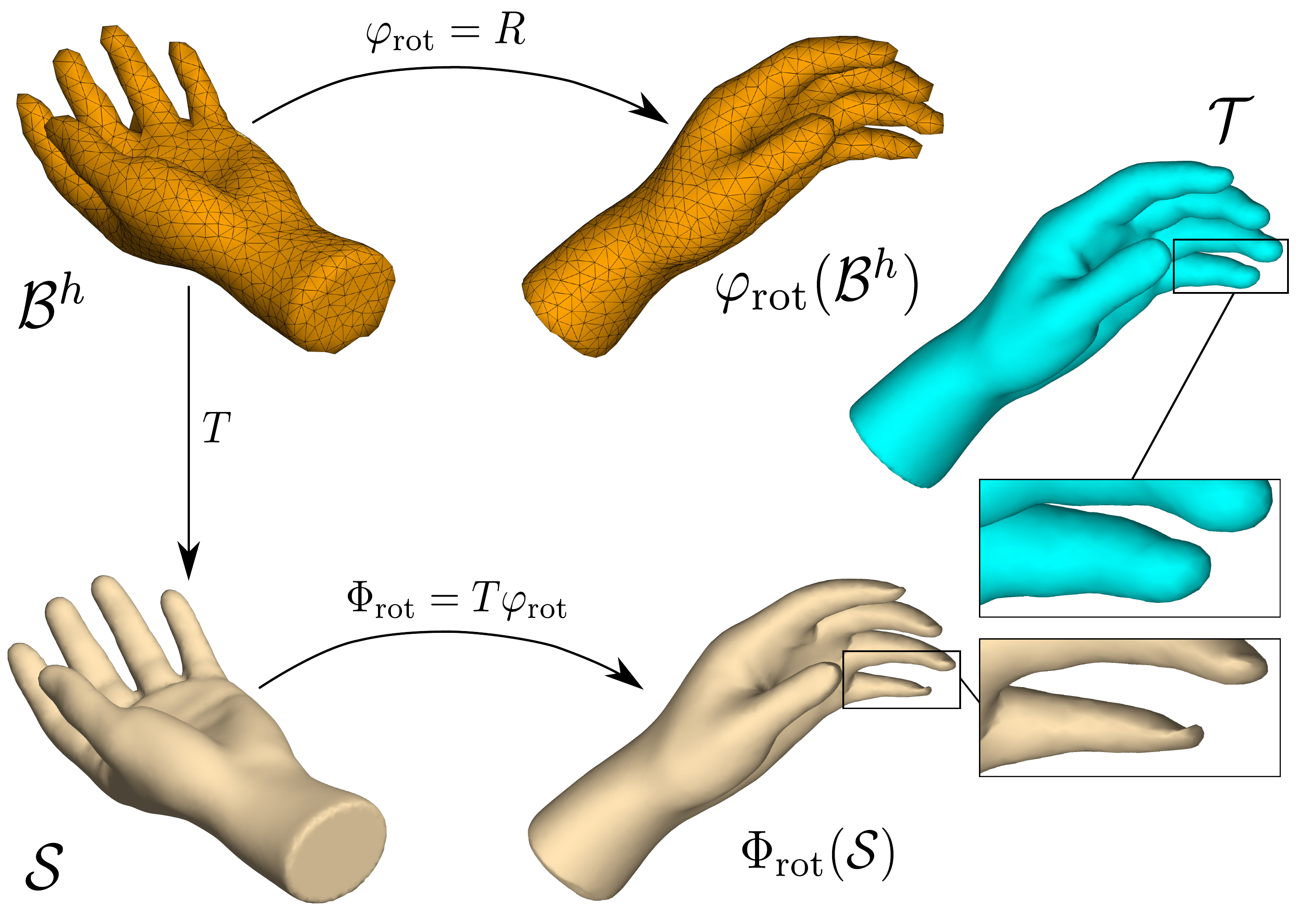}
  \caption{The effect of the interpolation of a pure rotation of the coarse
    mesh on the fine surface mesh. We chose a rotation $R\in SO(3)$ and
    constructed the map $T$ according to equation~(\ref{eq:4-2}). The rotated
    coarse mesh (orange) is shown in the upper row. The bottom row shows the
    effect of the interpolated rotation on the fine mesh. We compare this to
    the rotated fine mesh (turquoise).}
  \label{fig-8}
\end{figure}

Another drawback is due to the coarse volumetric meshes. For example, a pure
rotation of a coarse mesh does not necessarily induce a pure rotation of a
fine mesh. The reason for this is that we essentially map a linearly
interpolated map of the coarse mesh to the fine scale as described in
Section~\ref{s-4-1} without incorporating information in normal
direction. This effect is shown in Figure~\ref{fig-8}. Also, smoothing the
deformation after interpolation can smear features such as corners of the
deformed surface in an undesired way. Taking a finer volumetric mesh reduces
this effect.

Furthermore, the method is not guaranteed to avoid element flipping due to the
linearization and can exhibit high conformal element distortion, see
Figure~\ref{fig-7}. Methods dealing with this problem in two and three
dimensions can be found in~\cite{Kovalsky,Lipman3,Schuller} and can be
integrated into our framework.

Finally, the computed boundary forces that are responsible for the shape
change can be made invariant under rigid motion by a pull-back to the
undeformed configuration. Hence, we believe, that they can serve as a
comparison measure for different deformations of the same object since they
measure how ``difficult'' it is to achieve the observed change.

\vspace{0.5cm}

\textbf{Acknowledgments.}  This research was supported in part by the
U.S.-Israel Binational Science Foundation, Grant No. 2010331, by the Israel
Science Foundation, Grants No. 764/10 and 1265/14, by the Israel Ministry of
Science, and by the Minerva Foundation. The vision group at the Weizmann
Institute is supported in part by the Moross Laboratory for Vision Research
and Robotics.

\end{document}